\documentstyle[A4]{article}
\def\ko{ {\rm K}^ 0 }
\def\kob{ \overline{\rm K}{}^0 }
\def\kl{ {\rm K_{L}} }
\def\ks{ {\rm K_{S}} }
\def\gs{ {\it \Gamma}_{\rm S} }
\def\gl{ {\it \Gamma}_{\rm L} }
\def\kos{ | {\rm K^ 0} \rangle }
\def\kobs{ | {\rm \overline{K}{}^0} \rangle }
\def\be{ \begin{displaymath} }
\def\ee{ \end{displaymath} }
\def\ben{ \begin{equation} }
\def\een{ \end{equation} }
\def\bea{ \begin{eqnarray} }
\def\eea{ \end{eqnarray} }
\def\stimes{{\scriptstyle \times}}
\def\bo{ {\rm B}^ 0 }
\def\bob{ \overline{\rm B}{}^0 }
\def\bl{ {\rm B_{l}} }
\def\bh{ {\rm B_{h}} }
\begin{document}
\begin{titlepage}
\flushright{PSI-PR-93-18\\October 27, 1993}
\vspace{30mm}
\center{\Huge {\bf  Review on CP Violation} }
\vspace{20mm}
\center{\large Tatsuya Nakada\\[5mm]
Paul Scherrer Institute \\ CH-5232 Villigen-PSI, Switzerland}

\vspace{4cm}
\begin{abstract}
A phenomenological description of the neutral K and B meson systems is
presented. The situation of the current CP violation experiments is
described and a detailed discussion of their results is given,
followed by some future prospects.
\end{abstract}
\vspace{4cm}
\begin{center}
Plenary review talk given at XVI International Symposium on
Lepton-Photon Interactions at High Energies, August 10-15, 1993,
Cornell University, Ithaca.
\end{center}
\end{titlepage}
\newpage
\section{Introduction}
Symmetries are one of the most fundamental concepts in the laws of
nature leading to conserving quantities. Unexpected violations of
symmetries indicate some dynamical mechanism beyond the current
understanding of physics.

Observed P and C violations are well understood in the framework of
the standard model. They are naturally generated by left-handed
charged weak currents. On the other hand, the
origin of CP violation is still not explained. The standard electroweak
theory can accommodate CP violation with some complex elements in the
quark mass mixing matrix \cite{KM}. However, a possibility that CP
violation originates from some effect at a much higher energy scale is
not excluded \cite{super}. Interest in possible CPT violation has been
revived  by string theory \cite{superstring}.

Since the discovery of CP violating $\kl$ decays \cite{CPfirst}, the
neutral Kaon system is still the only system known to exhibit
violation of this symmetry. In this article, we recapitulate the
description of the neutral kaon system in order to outline necessary
experimental measurements which are still missing \cite{others}. It is
followed by a review on the current experimental situation and on
future prospects.

B-mesons appear to be the most attractive place to study CP violation
in a quantitative manner \cite{ali}. After a comparison of the K and B
meson systems, a short discussion of the experimental prospects in the
B sector is presented.

\section{Neutral K-Meson System}
\subsection{Basic Formalism}
Let $\kos$ and $\kobs$ be the stationary states of the
$\ko$-meson and its antiparticle $\kob$, respectively. Both
states are eigenstates of the strong and electromagnetic interaction
Hamiltonian, i.e.
\be
\left( H_{\rm st}+H_{\rm em} \right) \kos = m_0 \kos~~{\rm and}~~
\left( H_{\rm st}+H_{\rm em} \right) \kobs = \overline{m}{}_0 \kobs
\ee
where $m_0$ and $\overline{m}{}_0$ are the rest masses of $\ko$ and
$\kob$, respectively. The $\ko$ and $\kob$ states are connected through
CP transformations. For stationary states, T does not alter them with
the exception of an arbitrary phase. In summary, we obtain
\ben
\begin{array}{c}
C\!P \, \kos = e^{i \, \theta _{\rm CP} } \kobs~~{\rm and}~~
C\!P \, \kobs = e^{- \,i \, \theta_{\rm CP} } \kos~~ \\
T \, \kos = e^{ i \, \theta_{\rm T} } \kos~~{\rm and}~~
T \, \kobs = e^{ i \, \overline {\theta}{}_{\rm T } } \kobs
\end{array} \label{cp-phase}\een
where $\theta$'s are arbitrary phases and it follows that
\be
2 \, \theta_{\rm CP} =\overline{\theta}{}_{\rm T}\, - \, \theta_{\rm
T}~.
\ee
by assuming $C\!PT\, \kos = TC\!P \, \kos$.

If strong and electromagnetic interactions are
invariant under CPT transformation, which is assumed throughout this
paper, it follows that $m_0=\overline{m}{}_0$.

Next, we introduce a new interaction, $V$,
violates strangeness conservation. Through such interactions,
the K-mesons can decay into final states with no strangeness ($|\Delta
S|= 1$) and $\ko$ and $\kob$ can oscillate to each other
($|\Delta S| = 2$). Thus, a general
state $| \psi (t) \rangle$ which is a solution of the Schr\"odinger
equation
\ben
i \, \frac{\partial}{\partial t} |\psi (t)\rangle
 = \left( H_{\rm st} + H_{\rm em} + V \right) | \psi (t) \rangle
\label{schrodinger} \een
can be written as
\be
| \psi(t) \rangle = a(t) \kos + b(t) \kobs + \sum_{\rm f} c_{\rm f}
(t)| {\rm f} \rangle
\ee
where $a(t)$, $b(t)$ and $c_{\rm f}(t)$ are time dependent
functions. For a new interaction which is much weaker than strong and
electromagnetic
interactions, perturbation theory and the Wigner-Weisskopf
method \cite{wiger-weis} can be applied to solve equation
\ref{schrodinger}. We obtain
\ben
i \, \frac{ \partial }{\partial t}
\left( \begin{array}{c}a(t)\\b(t)\end{array}\right) =
\mbox{\boldmath$\Lambda$} \left(
\begin{array}{c}a(t)\\b(t)\end{array}\right)  = \left(
\mbox{\boldmath$M$} -
\, i \, \frac{\, \mbox{\boldmath$\Gamma$}\,}{2} \right) \left (
\begin{array}{c}a(t)\\b(t)\end{array}\right)
\label{basic}
\een
where the $2\times 2$ matrices \boldmath$M$\unboldmath and
\boldmath$\Gamma$\unboldmath are often referred to as the mass and
decay matrices.

The elements of the mass matrix are given as
\ben
M_{ij}= m_{0} \, \delta_{ij} + \langle i |V| j \rangle
+\sum_{\rm f}{\cal P}\left (\frac{\langle i|V|{\rm f} \rangle \langle
{\rm f} |V| j\rangle }{m_0-E_{\rm f}}\right ) \label{m12}
\een
where ${\cal P}$ stands for the principal part and the index $i=1$(2)
denotes $\ko$($\kob$). Let us split the Hamiltonian $V$ into the
known weak interaction part $H_{\rm weak}$ and a hypothetical
superweak interaction \cite{super},
$H_{\rm sw}$, i.e.
$V=H_{\rm weak}+H_{\rm sw}$. Since ordinary weak interactions do not
produce a direct $\ko$-$\kob$ transition, the second term of equation
\ref{m12} applies only for the superweak interaction for $i\neq j$. The
third term is dominated by the weak interaction since the second
order superweak interaction must be negligible. It follows that
\ben
M_{ij}= m_{0} \, \delta_{ij} + \langle i | H_{\rm sw} | j \rangle
+\sum_{\rm f}{\cal P}\left (\frac{\langle i| H_{\rm
weak}|{\rm f}\rangle
\langle {\rm f} |H_{\rm weak}| j\rangle }{m_0-E_{\rm f}}\right )~.
\label{m-element}
\een
Note that the sum is taken over {\it all possible intermediate
states} common to $\ko$ and $\kob$ for $i\neq j$.

The elements of the decay matrix are given by
\ben  {\it \Gamma}_{ij}=2\,
\pi\sum_{\rm f}\langle i|H_{\rm weak}|{\rm f} \rangle \langle {\rm
f}|H_{\rm weak}|j\rangle \delta(m_0-E_{\rm f})
\label{g-element}\een
The sum is taken over only {\it real final states} common to $\ko$ and
$\kob$ for $i\neq j$. Since ${\it \Gamma}_{ij}$ starts from second
order, the superweak Hamiltonian can be neglected.

If Hamiltonians are not Hermitian, transition probabilities are not
conserved in decays or oscillations, i.e. the number of initial
particles is not identical to the number of final particles. This is
also referred as break down of unitarity. {\bf We assume from now
on that all the Hamiltonians are Hermitian}.

If $V$ is Hermitian and invariant under T, CPT or CP transformations,
the mass and decay matrices must satisfy the following conditions;
\bea
{\rm T}~&:&~~\left| M_{12}-i\, \frac{ {\it \Gamma}_{12} }{2} \right|
= \left| M_{12}^* -i\, \frac{ {\it \Gamma}_{12}^* }{2} \right|
\nonumber
\\ {\rm CPT}~&:&~~M_{11}=M_{22},~{\it \Gamma}_{11}={\it \Gamma}_{22}
\nonumber \\
{\rm CP}~&:&~~\left| M_{12}-i\, \frac{ {\it \Gamma}_{12} }{2}
\right| = \left| M_{12}^*-i\, \frac{ {\it \Gamma}_{12}^* }{2} \right|,~
M_{11}=M_{22},~{\it \Gamma}_{11}={\it \Gamma}_{22}
\nonumber
\eea
where equations \ref{cp-phase}, \ref{m-element} and \ref{g-element}
are used. It follows that
\ben
\begin{array}{l}
\bullet {\rm if}~M_{11} \neq M_{22}~{\rm or}~ {\it \Gamma}_{11}\neq
{\it
\Gamma}_{22}~:\\
{}~~{\rm{ \bf CPT~and~CP}~are~violated}
\vspace{2 mm}
\\
\bullet {\rm if}~\sin \left(\varphi_{\it \Gamma} -
\varphi_{M}\right) \neq 0~:\\ ~~
{\rm {\bf T}~(or~unitarity)~{\bf and~CP}~are~violated}~.
\label{CPCPT}
\end{array}
\een
where $\varphi_{M}=\arg\left( M_{12} \right)~~{\rm and}~~ \varphi_{\it
\Gamma}=\arg\left( {\it \Gamma}_{12}\right )$.

Note that CP is not conserved in both above cases; i.e. CP
violation in the mass and decay matrices cannot be separated from CPT
violation or T violation.

Solutions of equation \ref{basic} for initially pure $\ko$ and
$\kob$ states are given by
\bea |{\rm K}{}^0(t)\rangle &=& \left[\, f_+(t) - 2\,
\varepsilon_{\rm CPT}
\,f_-(t)\, \right] \kos  + \left(\, 1 - 2\, \varepsilon_{\rm T}
\, \right)\, e^{-\,i\,
\varphi_{\it \Gamma}} f_-(t)\, \kobs \label{k0state}\\
&=&\frac{1}{\,\sqrt{\, 2 \,}\,} \left(\, |\ks
\rangle\, e^{-\, i\,
\lambda_{\rm S}\, t } + |\kl \rangle\, e^{-\, i\, \lambda_{\rm
L}\, t }\,
\right)
\label{initialk0} \eea
and
\bea
|\overline{\rm K}{}^0(t)\rangle &=& \left(\, 1 + 2\,
\varepsilon_{\rm T}
\, \right) \, e^{i\,\varphi_{\it \Gamma}}\,f_-(t) \,
\kos  + \left[\, f_+(t) + 2\, \varepsilon_{\rm CPT} \,f_-(t)\, \right]
\, \kobs \label{k0bstate}\\
&=&\frac{\, 1 + 2\,
\varepsilon_{\rm T}
\,}{\,\sqrt{\, 2 \,}\,} \, e^{i\,\varphi_{\it \Gamma}}\nonumber \\
&&~~~\times \left[\,\left( 1+2\, \varepsilon_{\rm CPT} \right) |\ks
\rangle\, e^{-\, i\,
\lambda_{\rm S}\, t } -\left( 1-2\, \varepsilon_{\rm CPT}
\right)|\kl \rangle\, e^{-\, i\, \lambda_{\rm L}\, t }\,
\right]
\nonumber \eea
where
\be
f_{\pm}(t)=\frac{1}{\,2\,}\left(\, e^{-\, i\, \lambda_{\rm S}\, t }
\pm
 e^{-\, i\, \lambda_{\rm L}\, t}\, \right)~.
\ee
The parameters $\lambda_{\rm S}$ and $\lambda_{\rm L}$ are eigenvalues
of $\mbox{\boldmath$\Lambda$}$, and $\ks$ and $\kl$ are the
corresponding eigenstates given by
\ben \begin{array}{lcl}
|\ks \rangle &=& \frac{\displaystyle 1}{\displaystyle \, \sqrt{\,2\,}
\,}\left[
\left(\, 1\,-\,2\, \varepsilon_{\rm CPT}\, \right) \kos \, + \,
\left(\, 1\,-\,2\, \varepsilon_{\rm T}\, \right)\, e^{-\,
i\,\varphi_{\it \Gamma}}\,
\kobs
\,
\right]
\vspace{2mm} \\
|\kl \rangle &=& \frac{\displaystyle 1}{\displaystyle \, \sqrt{\,2\,}
\,}\left[
\left(\, 1\,+\,2\, \varepsilon_{\rm CPT}\, \right) \kos \, - \,
\left(\, 1\,-\,2\, \varepsilon_{\rm T}\, \right)\, e^{-\,
i\,\varphi_{\it
\Gamma}} \kobs \, \right] ~.
\end{array}\label{klks}\een
They have definite masses and decay widths given by
$\lambda_{\rm S}$ and $\lambda_{\rm L}$ as
\be
\lambda_{\rm S(L)}= m_{\rm S(L)}\, - \,i\, \frac{ \, {\it
\Gamma}_{\rm S(L)}
\, }{ 2}
\nonumber \ee
with
\bea
m_{\rm S(L)}&=& \frac{\, M_{11}+M_{22}\,}{2} \,
+\!(-)\, \Re  \left( \sqrt{\, {\it \Lambda}_{12}\,{\it
\Lambda}_{21}\,}\, \right) \nonumber \\
&=&\frac{\, M_{11}+M_{22}\,}{2} \, -\!(+)\,\left| M_{12} \right|
\nonumber \eea
and
\bea {\it \Gamma}_{\rm S(L)}&=& \frac{\, {\it
\Gamma}_{11}+{\it \Gamma}_{22}\,}{2}
\, -\!(+)\,2\,
\Im
\left(
\sqrt{\, {\it \Lambda}_{12}\,{\it
\Lambda}_{21}\,}\, \right) \nonumber \\ &=&\frac{\,
{\it \Gamma}_{11}+{\it \Gamma}_{22}\,}{2}
\, +\!(-)\,\left| {\it \Gamma}_{12} \right|
\nonumber \eea
where we used
\be
\varphi_{\it \Gamma} - \varphi_{M} = \pi \, -\, \delta  \varphi,~~
\left| \delta  \varphi \right| \ll 1
\ee
and
\be
\left| {\it \Lambda}_{22} \, -\, {\it \Lambda}_{11} \right| \ll 1
\ee
which are derived from empirical facts,
$m_{\rm L}>m_{\rm S}$,
$\gs>\gl$ and small CP violation. Note that the current values
given by Particle Data Group \cite{ref_pdg} are
\be
\begin{array}{lcl}
\Delta m &\equiv& m_{\rm L} - m_{\rm S} = 2\, \left| M_{12} \right|
= (0.5351\pm0.0024)\stimes10^{10}~\hbar {\rm s}^{-1}
\vspace{2 mm}\\
\Delta {\it \Gamma} &\equiv& {\it \Gamma}_{\rm S} - {\it \Gamma}_{\rm
L}= 2 \, \left| {\it \Gamma}_{12} \right|
=(1.1189\pm0.0025)\stimes 10^{10}~ {\rm s}^{-1}~.
\end{array}\ee

The two CP violation parameters $\varepsilon_{\rm T}$ and
$\varepsilon_{\rm CPT}$ are given by
\bea
\varepsilon_{\rm T}&=&\frac{ \Delta m\, \Delta {\it \Gamma} }{\, 4\,
\Delta m^2\,+\, \Delta {\it \Gamma}^2\,} \left( 1\, + \,i\, \frac{\,
2\,
\Delta m \,}{\Delta {\it \Gamma}} \right) \, \delta \varphi
\nonumber \\
\varepsilon_{\rm CPT}&=&\frac{i\,2\, \Delta {\it \Gamma} }{\, 4\,
\Delta m^2\,+\, \Delta {\it \Gamma}^2\,} \left( 1\, + \,i\, \frac{\,
2\,
\Delta m \,}{\Delta {\it \Gamma}} \right) \, \left({\it
\Lambda}_{22}\, - \,{\it \Lambda}_{11}\right)~.
\nonumber \eea
As seen from the statements \ref{CPCPT}, $\varepsilon_{\rm T}\neq 0$
implies CP and T violation, and $\varepsilon_{\rm CPT}\neq 0$ means CP
and CPT violation. It should be noted that both $\varepsilon_{\rm T}$
and
$\varepsilon_{\rm CPT}$ {\bf do not depend on any phase convention}.
The phase of $\varepsilon_{\rm T}$ is given by the $\ks$-$\kl$ mass and
decay width differences which are not related to CP violation. This
phase is often referred to as ``superweak'' phase:
\be
\phi_{\rm sw}=\arg\left( \varepsilon_{\rm T}\right) = \tan
^{-1}\left(\frac{\, 2\, \Delta m \,}{\Delta {\it \Gamma} }\right)~.
\ee

If we assume that ordinary weak interactions conserve CPT, i.e. ${\it
\Gamma}_{11}={\it \Gamma}_{22}$, the phase of the CP and CPT violation
parameter $\varepsilon_{\rm CPT}$ is given by
\be
\arg \left( \varepsilon_{\rm CPT} \right) = \phi_{\rm sw}+\frac{\,
\pi \,}{2}~.
\ee

\subsection{CP Violation and Semileptonic Decays}
The instantaneous decay amplitudes for $\ko$ and $\kob \rightarrow \ell
^+ \pi^- \nu$ are given by
\be
A_+=\langle \pi^-(\vec{p}_{\pi}), \ell^+ (\vec{p}_{\ell},
\vec{s}), \nu(\vec{p}_{\nu}) | H_{\rm weak}\kos
\ee
and
\be \overline{A}_+=\langle  \pi^-(\vec{p}_{\pi}), \ell^+
(\vec{p}_{\ell},
\vec{s}), \nu(\vec{p}_{\nu}) |
H_{\rm weak}\kobs~.
\ee
Parameters $\vec{p}$ and $\vec{s}$ are the
momentum vectors and spin, respectively. In the standard model,
$\overline{A}_+$ is very strongly suppressed compared with $A_+$
($\Delta Q=\Delta S$ rule). Similarly, we can define decay amplitudes
for $\ko$ and $\kob \rightarrow \ell ^- \pi^+ \overline{\nu}$ to be
\be A_-=\langle \pi^+(\vec{p}_{\pi}), \ell^- (\vec{p}_{\ell},
-\,\vec{s}), \overline{\nu}(\vec{p}_{\nu}) | H_{\rm weak}\kos
\ee
and
\be \overline{A}{}_-=\langle  \pi^+(\vec{p}_{\pi}), \ell^-
(\vec{p}_{\ell},
-\,\vec{s}), \overline{\nu}(\vec{p}_{\nu}) | H_{\rm weak}\kobs~.
\ee
Note that the spin of the lepton is reversed. The decay amplitude
$A_{-}$ violates the $\Delta Q= \Delta S$ rule.

If we assume that CPT is conserved in the ordinary weak interaction,
the following relations can be obtained;
\be
\left|A_+\right|=\left|
\overline{A}{}_-\right|,~\left|A_-\right|=\left| \overline{A}_+\right|
\ee
and
\be
A_+\,\overline{A}{}_+^* = A_-\, \overline{A}{}_-^*~.
\ee
These relations do not depend on any phase convention.

The parameter $x$ defined as
\ben
 x=\frac{\sum_{\vec{s}}\int {\rm d}\Omega\, A_+^*\,
\overline{A}{}_+\,e^{-\, i\,\varphi_{\it
\Gamma}}}{\sum_{\vec{s}}\int {\rm d}\Omega \,
\left|A_+\right|^2}
= \frac{\sum_{\vec{s}}\int {\rm d}\Omega\, A_-^*\,
\overline{A}{}_-\,e^{-\, i\,\varphi_{\it
\Gamma}}}{\sum_{\vec{s}}\int {\rm d}\Omega \,
\left|\overline{A}{}_-\right|^2}
\label{xparam}\een
is a measure for the violation of $\Delta Q = \Delta S$. Note that $x$
{\bf does not depend on any phase convention}.

Using equations \ref{klks}, the charge asymmetry in the semileptonic
$\kl$ decay is now given as
\bea
\delta_{\ell}&=&\frac{\, {\it \Gamma}(\kl \rightarrow \pi^- \ell^+
\nu)\,-\,{\it \Gamma}(\kl \rightarrow \pi^+ \ell^-
\overline{\nu}) \,}{\, {\it \Gamma}(\kl \rightarrow \pi^- \ell^+
\nu)\,+\,{\it \Gamma}(\kl \rightarrow \pi^+ \ell^-
\overline{\nu}) \,} \nonumber \\
&=& 2\, \left[ \Re\left(\varepsilon_{\rm CPT}\right) +
\Re\left(\varepsilon_{\rm T}\right) \right] .
\nonumber \eea
where ${\cal O}(\varepsilon_{\rm CPT})\approx {\cal
O}(\varepsilon_{\rm T})\approx x \ll 1$ is assumed. It has to be noted
that the experimentally well established CP violation
effect \cite{ref_pdg}
\be
\delta_{\ell}=\left(0.327\pm 0.012\right) \stimes 10^{-2}
\ee
cannot tell whether it is violation of CP and CPT or CP and T.

The separation of CP and CPT violating processes from CP and T
violating processes can be done if we start from the initially pure
$\ko$ and $\kob$ states \cite{cabbiar}. Let us consider the following
four time dependent decay rates:
\begin{itemize}
\item $R_+(t)$:\hspace{5mm}
$\ko$ at $t=0$ decaying into $\ell ^+ \pi^- \nu$ at time $t$
\item $R_-(t)$:\hspace{5mm}
$\ko$ at $t=0$ decaying into $\ell ^- \pi^+ \overline{\nu}$ at time $t$
\item $\overline{R}{}_+(t)$:\hspace{5mm}
$\kob$ at $t=0$ decaying into $\ell ^+ \pi^- \nu$ at time $t$
\item $\overline{R}{}_-(t)$:\hspace{5mm}
$\kob$ at $t=0$ decaying into $\ell ^- \pi^+ \overline{\nu}$ at time
$t$.
\end{itemize}
These four rates can be obtained using equations
\ref{k0state} and \ref{k0bstate} as
\bea
R_+(t)~\left(\overline{R}_-(t) \right)&\propto&
\left[ 1 -\!(+)\, 4\, \Re(\varepsilon_{\rm CPT}) + 2\, \Re(x) \right]
e^{-\, i\, \gs \, t}\nonumber \\
&&~~
+ \left[ 1 +\!(-)\, 4\, \Re(\varepsilon_{\rm CPT}) - 2\, \Re(x)
\right] e^{-\, i\, \gl \, t}
\nonumber \\ &&~~~~
+ 2\, e^{-\, {\overline{\it \Gamma}}\, t} \cos ( \Delta m \, t)
\nonumber \\ &&~~~~~
+\!(-)\, 4 \left[ 2\, \Im (\varepsilon_{\rm CPT}) - \Im (x) \right]
e^{-\, {\overline{\it \Gamma}}\, t} \sin ( \Delta m \, t)
\nonumber \eea
and
\bea
R_-(t)~\left(\overline{R}_+(t) \right)&\propto&
\left[ 1 -\!(+)\, 4\, \Re(\varepsilon_{\rm T}) + 2\, \Re(x) \right]
e^{-\, i\, \gs \, t}\nonumber \\ &&~~
+ \left[ 1 -\!(+)\, 4\,
\Re(\varepsilon_{\rm T}) - 2\, \Re(x)
\right] e^{-\, i\, \gl \, t}
\nonumber \\ &&~~~~
- 2\left[ 1 -\!(+)\, 4\, \Re(\varepsilon_{\rm T})\right] e^{-\,
{\overline{\it \Gamma}}\, t} \cos ( \Delta m \, t)
\nonumber \\ &&~~~~~ -\!(+)\, 4  \Im (x) \,  e^{-\,
{\overline{\it \Gamma}}\, t} \sin ( \Delta m
\, t)
\nonumber
\eea

For simplicity, let us assume that the $\Delta Q=\Delta S$ rule
holds for the moment. Then, $R_+(t)$ and $\overline{R}{}_-(t)$ are due
to the processes where $\ko$ remains $\ko$ and $\kob$ remains $\kob$,
respectively. The two processes are CP and CPT conjugate to each other.
Therefore, $R_+(t)\neq \overline{R}{}_-(t)$ must be a sign of CP and
CPT violation. Indeed, a time dependent CP asymmetry $A_{\rm
CPT}(t)$ given as
\bea
A_{\rm CPT}(t)&=&\frac{\,\overline{R}{}_-(t) -
R_+(t)\,}{\,\overline{R}{}_-(t) +  R_+(t)\,} \nonumber \\
&=&4 \frac{\,\Re\left(\varepsilon_{\rm CPT}\right)\, \left(e^{-\,\gs
\,t} - e^{-\, \gl\, t} \right) -
\Im\left(2\, \varepsilon_{\rm CPT}- x \right)
e^{-\, \overline{\it \Gamma}\, t}
\sin \left(\Delta m \, t\right)\, }
{\,e^{-\,\gs \,t} + e^{-\, \gl\, t} +2\, e^{-\,
\overline{\it \Gamma}\, t} \cos \left(\Delta m
\, t\right)\,}
\nonumber \eea
depends only on $\varepsilon_{\rm CPT}$ but not on $\varepsilon_{\rm
T}$. Note that $\overline{\it
\Gamma}$ is the average decay width $(\gl +\gs)/2$. For a large decay
time
$t$, we obtain
\be
\lim_{t\rightarrow \infty} A_{\rm CPT}(t) = - 4\,
\Re\left(\varepsilon_{\rm CPT}\right)~.
\ee
$R_-(t)$ and $\overline{R}{}_+(t)$ are due to oscillations of $\ko$
into
$\kob$ and $\kob$ into $\ko$ respectively, which are CP and T conjugate
to each other. The CP asymmetry defined as
\bea
A_{\rm T}(t)&=&\frac{\,\overline{R}{}_+(t) -
R_-(t)\,}{\,\overline{R}{}_+(t) +  R_-(t)\,} \nonumber \\
&=&4\, \Re \left( \varepsilon_{\rm T} \right)
- \frac{4\, \Im \left(x \right) e^{-\, \overline{\it
\Gamma}\, t} \sin \left(\Delta m \, t\right) }
{\,e^{-\,\gs \,t} + e^{-\, \gl\, t} -2\, e^{-\,
\overline{\it \Gamma}\, t} \cos \left(\Delta m
\, t\right)\,}
\nonumber \eea
is sensitive only to CP and T violation with
\be
\lim_{t\rightarrow \infty} A_{\rm T}(t) = 4\,
\Re\left(\varepsilon_{\rm T}\right)~.
\ee

It must be noted that $A_{\rm T}$ depends only on the
real part of $\varepsilon_{\rm T}$ if the $\Delta Q= \Delta S$ rule is
valid. We conclude that CP violation in oscillations which is
generated by the interference between the mass and decay matrices
contributes only to the {\bf real part} of $\varepsilon_{\rm T}$.

\subsection{CP Violation and Two-Pion Decays}
Two-pion final states are common to both $\ko$ and $\kob$ decays.
Since they must be eigenstates of electromagnetic and strong
interactions, we consider them to be in the isospin eigenstates. The
total angular momentum of the two-pion final state is 0 and only
isospin states $I=0$ and $I=2$ are allowed due to Bose
statistics. The amplitude of a $\ko$ decaying into $2\pi(I=0)$ is given
by
\be
A_{I=0}={}_{\rm out} \langle 2\pi(I=0)|\, H_{\rm weak} \, \kos
\ee
where $| 2\pi(I=0) \rangle_{\rm out}$ denotes the state for out-going
two-pions with $I=0$. By assuming that the weak interaction is
invariant under CPT transformation and that the S-matrix for the
two-pion state is unitary and diagonal, we can relate $A_{I=0}$ to
the $\kob$ decay amplitude by
\be
\overline{A}{}_{I=0}={}_{\rm out} \langle 2\pi(I=0)|\, H_{\rm
weak}
\,
\kobs=A_{I=0}^* \, e^{i\, 2 \delta_0}  e^{i \, \left(
\theta_{\rm CP}- \overline{\theta}{}_{\rm T}\right)}
\ee
where $\delta_{0}$ is the strong interaction phase shift measured in
the $\pi$-$\pi(I=0)$ elastic scattering at $\sqrt{s}=m_{\ko}$.
The arbitrary phases, $\theta$'s are defined in equations
\ref{cp-phase}. It is common to write
\ben \begin{array}{lcl}
A_{I=0}&=&a_0\, e^{i\,\delta_0} \vspace{2mm} \\
\overline{A}{}_{I=0}&=&a_0^*\, e^{i\,\left( \delta_0 + \theta_{\rm
CP}-\overline{\theta}{}_{\rm T}\right)}
\end{array} \label{amplitude} \een
where $a_0$ contains only the weak interaction part. Similarly for
the $2\pi(I=2)$ state, we have
\be \begin{array}{lcl}
A_{I=2}=a_2\, e^{i\,\delta_2} \vspace{2mm} \\
\overline{A}{}_{I=2}=a_2^*\, e^{i\,\left( \delta_2 + \theta_{\rm
CP}-\overline{\theta}{}_{\rm T}\right)}~.
\end{array} \ee
Experimentally we have \cite{devlin}
\be
\left| \frac{\, a_2 \,}{\, a_0 \,} \right|\approx 0.045
\ee
i.e. the decay into the $I=2$ state is suppressed ($\Delta
I=1/2$ rule).

Physical two-pion states are $\pi^+ \pi^-$ and $\pi^0 \pi^0$. The
instantaneous $\ko$ decay width into $\pi^+ \pi^-$ is given by
\be
{\it \Gamma}(\ko \rightarrow \pi^+ \pi^-) =
\left| \sqrt{\, \frac{\, 2\,}{\, 3\,}\,} \, A_{I=0} +
\sqrt{\, \frac{\, 1\,}{\, 3\,}\,} \, A_{I=2} \right|^2
\ee
Since $\ko \rightarrow \pi^+ \pi^-$ and $\kob \rightarrow \pi^+ \pi^-$
are CP conjugate processes, the ratio
\bea
\frac{\,{\it \Gamma}(\kob \rightarrow \pi^+ \pi^-) -
{\it \Gamma}(\ko \rightarrow \pi^+ \pi^-) }
{\,{\it \Gamma}(\kob \rightarrow \pi^+ \pi^-) + {\it \Gamma}(\ko
\rightarrow \pi^+ \pi^-) } &=& \sqrt{\, 2\,} \, \Im \left(
\frac{\, a_2\,}{\, a_0\,}\right) \, \sin \left(\delta_2 - \delta_0
\right) \nonumber \\
&\equiv& - 2\, \Re \left( \varepsilon ' \right) \label{eppp}
\eea
where
\be
\varepsilon' = \frac{\, i \,}{\, \sqrt{2}\,} \,
\Im \left(
\frac{\, a_2\,}{\, a_0\,}\right)\, e^{i \, \left(\delta_2 - \delta_0
\right)}
\ee
is a measure of CP violation. A non-zero value of $\Re
\left(\varepsilon' \right)$ implies CP violation in the decay amplitude
($|\Delta S|=1$) which is often referred to as ``direct CP violation''
. This requires that the strong interaction phase shifts for $I=0$ and
$I=2$ states and the phases of $I=0$ and $I=2$ weak decay amplitudes
are {\bf both} different. The former is experimentally measured to be
\cite{ox}
\be
\delta_0 - \delta_2 = (47\pm 6)^\circ~,
\ee
If CP violation is due to the superweak interaction $H_{\rm sw}$, no CP
violation is expected in the decay amplitudes and $\varepsilon'=0$. On
the other hand, we do expect some phase difference between the $I=0$
and $I=2$ weak decay amplitudes if CP violation is due to the weak
interaction.

It must be noted again that CP violation in the decay amplitude, which
is generated by the interference between two decay amplitudes
contributing to the same final state, depends only on the {\bf real
part} of $\varepsilon '$ as seen from equation
\ref{eppp}.

Since $\ko$ and $\kob$ are not mass eigenstates, it is
more suitable to discuss $\ks$ and $\kl$ decays. The two-pion final
states are CP eigenstates with an eigenvalue of $+1$. In the limit of
CP conservation, we have $\varepsilon_{\rm CPT}=0$, $\varepsilon_{\rm
T}=0
$ and $\varphi_{\it \Gamma}=\theta_{\rm CP}$ where $\theta_{\rm CP}$
is an arbitrary CP phase of the neutral kaon system defined in
equations \ref{cp-phase}. In this limit, the mass eigenstates $\ks$
and $\kl$ become also CP eigenstates with eigenvalues $+1$ and
$-1$ respectively. Then, a parameter defined as
\be
\eta_{I=0} = \frac{\, \langle 2\, \pi(I=0) \left| H_{\rm weak}  \right|
\kl
\rangle \,}{\, \langle 2\, \pi(I=0) \left| H_{\rm weak}  \right| \ks
\rangle \,}
\ee
is a measure of CP violation. Using equations
\ref{klks} and \ref{amplitude}, it follows that
\ben
\eta_{I=0}=\varepsilon_{\rm CPT} + \varepsilon_{\rm T} +
\frac{\,i\,}{\,2\,}
\left( 2\,\varphi_0 + \varphi_{\it \Gamma}+\overline{\theta}{}_{\rm
T}-\theta_{\rm CP}  \right)
\label{eta0}\een
where $\varphi_0=\arg \left(a_0\right)$. Note that both
$\varepsilon_{\rm CPT}$ and $\varepsilon_{\rm T}$ contribute to
$\eta_{I=0}$. The third term is usually considered to be $0$. This
will be examined in detail later. A commonly used parameter
$\varepsilon$, which is somewhat ambiguous due to the phase convention,
is here given by
\be
\varepsilon= \varepsilon_{\rm T} +
\frac{\,i\,}{\,2\,}
\left( 2\,\varphi_0 + \varphi_{\it \Gamma}+\overline{\theta}{}_{\rm
T}-\theta_{\rm CP}  \right)~.
\ee

The {\bf imaginary part} of $\eta_{I=0}$ is interpreted as CP violation
generated by the interference between oscillations and decays.

The CP violation parameters for the $I=2$ state is given by
\be
\eta_{I=2}=\varepsilon_{\rm CPT} + \varepsilon_{\rm T} +
\frac{\,i\,}{\,2\,}
\left( 2\,\varphi_2 + \varphi_{\it \Gamma}+\overline{\theta}{}_{\rm
T}-\theta_{\rm CP}  \right)
\ee
where $\varphi_2=\arg \left(a_2\right)$. If CP violation in
the decay amplitude is present, $\varphi_0\neq
\varphi_2$ thus we expect $\Im\left(
\eta_{I=0}\right) \neq \Im \left( \eta_{I=2} \right)$.

CP violation parameters measured directly in the experiment are
\bea
\eta_{+\,-} &=& \frac{\,\langle \pi^+ \pi^-  \left| H_{\rm weak}
\right|
\kl
\rangle \,}{\, \langle \pi^+ \pi^- \left| H_{\rm weak}  \right| \ks
\rangle \,}
\nonumber \\
&=& \eta_{I=0} + \varepsilon '
\label{reta+-} \eea
and
\bea
\eta_{0\,0} &=& \frac{\,\langle \pi^0 \pi^0  \left| H_{\rm weak}
\right|
\kl
\rangle \,}{\, \langle \pi^0 \pi^0 \left| H_{\rm weak}  \right| \ks
\rangle \,}
\nonumber \\ &=& \eta_{I=0} - 2\, \varepsilon '~.
\label{reta00} \eea
With the presence of direct CP violation, we expect $\eta_{+\, -} \neq
\eta_{0\, 0}$.

Since $\varepsilon '$ is measured to be very small and
$\arg(\varepsilon ')\approx \arg(\varepsilon_{\rm T})\approx
45^\circ$, the phase of
$\eta_{+ \, -}$, $\phi_{+\, -}$ is essentially identical to the phase
of
$\eta_{I=0}$, $\phi_{0\, 0}$.

\subsection{CP Violation and Three-Pion Decays}
The CP eigenvalue of the three-pion final state depends on the angular
momentum configuration. For $\pi^+ \pi^- \pi^0$ mode, we denote
$l_{\pi^+ \pi^-}$ to be the angular momentum between $\pi^+$ and
$\pi^-$. The CP eigenvalue of the $\pi^+ \pi^-$ pair is always $+1$.
Then the CP eigenvalue of the
$\pi^+ \pi^- \pi^0$ system is given by $-(-1)^{l'}$ where $l'$ is the
relative angular momentum between the
$\pi^+ \pi^-$ pair and the remaining $\pi^0$ (see figure
\ref{fig-3pi}). Due to conservation of angular momentum, we have
$l_{\pi^+ \pi^-}=l'$ for the neutral K-meson final state: i.e.
\be
CP(\pi^+ \pi^- \pi^0 ) = - (-1)^{l_{\pi^+ \pi^-}}.
\ee
Since the kaon mass is very close to the three-pion mass, higher
angular momentum states are suppressed due to the centrifugal barrier.
Therefore, we consider only
$l_{\pi^+ \pi^-}=0$ or 1. Note that the angular momentum
part of the wave function is symmetric for the $l_{\pi^+ \pi^-}=0$
state and antisymmetric for the $l_{\pi^+ \pi^-}=1$ state over the
exchange of $\pi^+$ and $\pi^-$.
\begin{table}[b]
\begin{minipage}[t]{60mm}
\vspace{35mm}
%\special{picture 3pi.pict}
\special{illustration 3¹-configuration.eps scaled 1200}
\refstepcounter{figure}
\label{fig-3pi}  {\bf FIG. \thefigure.} Angular momentum configuration
of the $\pi^+
\pi^- \pi^0$ final state.
\end{minipage}
\begin{minipage}[t]{5mm}
\hspace{5mm}
\end{minipage}
\hspace{5mm}
\begin{minipage}[t]{70mm}
\caption{Allowed isospin and angular momentum configurations for the
$\pi^+ \pi^- \pi^0$ system. S and AS stand for symmetric and
antisymmetric.}
\label{tab_isospin}
\begin{tabular}{lcccc}
$CP(\pi^+ \pi^- \pi^0)$
&\multicolumn{2}{c}{$+1$}&\multicolumn{2}{c}{$-1$}\\
\hline
$ l_{\pi^+ \pi^-}$&\multicolumn{2}{c}{$1$}&\multicolumn{2}{c}{$0$}\\
$l$-Sym.($\pi^+\leftrightarrow
\pi^-$)&\multicolumn{2}{c}{AS}&\multicolumn{2}{c}{S} \\
$I_{3\pi}$& 0 & 2 & 1 & 3 \\
$ I_{\pi^+ \pi^-}$& 1 & 1 & 0 or 2& 2 \\
$I$-Sym.($\pi^+\leftrightarrow \pi^-)$ & AS & AS & S & S \\
\end{tabular}
\end{minipage}
\end{table}

The $\pi^+$$\pi^-$ subsystem can have an isospin of
$I_{\pi^+\pi^-}=0$, 1 or 2. The isospin part of the wave function is
symmetric for $I_{\pi^+\pi^-}=0$ and 2 over the exchange of $\pi^+$
and $\pi^-$. The $I_{\pi^+\pi^-}=1$ state is antisymmetric. The total
isospin of the $\pi^+ \pi^- \pi^0$ system is $I_{\rm 3\pi}= 0$, 1,  2
or 3 where $I_{\pi^+\pi^-}=0$ contributes to $I_{3\pi}=1$,
$I_{\pi^+\pi^-}=1$ to $I_{3\pi}=0$ and 2 and $I_{\pi^+\pi^-}=2$ to
$I_{3\pi}=3$.

Since the total wave function must be symmetric
over the exchange of $\pi^+$ and $\pi^-$,
the $l_{\pi^+ \pi^-}=1$ state has $I_{3\pi}=0$ or 2 and the
$l_{\pi^+
\pi^-}=0$ state $I_{3\pi}=1$ or 3. Table \ref{tab_isospin}
summarises  the allowed $\pi^+ \pi^- \pi^0$ configuration

The $I_{3\pi}=3$ state is expected to be suppressed compared with
$I_{3\pi}=1$ state due to the $\Delta I=1/2$ rule. The $I_{3\pi}=0$
state is suppressed since the isospin part of the wave function must
be totally antisymmetric over the exchange of $\pi^+$, $\pi^-$ and
$\pi^0$ which requires that it must be at least in the third order of
the pion kinetic energies \cite{zemach}. In conclusion, we have
\begin{itemize}
\item $\pi^+ \pi^- \pi^0 (CP=-1)$\\
$l_{\pi^+ \pi^-}=l' = 0$, $I_{3\pi}=1$
\item $\pi^+ \pi^- \pi^0 (CP=+1)$\\
$l_{\pi^+ \pi^-}=l'=1$, $I_{3\pi}=2$
\end{itemize}

Using the invariant masses of $\pi^+ \pi^0$ and
$\pi^- \pi^0$ pairs denoted by $m_{+\, 0}$ and $m_{-\, 0}$
respectively, the decay amplitudes are given as
\bea
A(\ko \rightarrow \pi^+ \pi^- \pi^0 )   &=&
a(I_{3\pi}\!=\!2,m_{+\, 0},m_{-\, 0})\, e^{i\,
\delta_{3 \pi(I=2)}} \nonumber \\
&&~~~~+ a(I_{3\pi}\!=\!1,m_{+\, 0},m_{-\,
0})\, e^{i\, \delta_{3 \pi(I=1)}}
\label{3pi1} \eea
and
\bea A(\kob \rightarrow \pi^+ \pi^- \pi^0 )   &=&
a^*(I_{3\pi}\!=\!2,m_{+\, 0},m_{-\, 0})\, e^{i\,
\left( \delta_{3 \pi\,(I=2)}+\theta_{\rm CP} -
\overline{\theta}{}_{\rm T}\right)}
\label{3pi2}\nonumber
\\  && - a^*(I_{3\pi}\!=\!1,m_{+\,
0},m_{-\, 0})\, e^{i\,\left( \delta_{3 \pi\,(I=1)}+\theta_{\rm CP} -
\overline{\theta}{}_{\rm T}\right)}
\eea
where the $\delta$'s are strong interaction phase shifts for the
corresponding isospin states. Note that the phase space integrations
give
\be
\int\limits_{m_{+\, 0}>m_{-\,
0}}\!\!\!\!\!\!\! {\rm d} \Omega \,
a(I_{3\pi}\!=\!2,m_{+\, 0},m_{-\,
0})= -\!\!\!\!\!\!\!\!\!\! \int\limits_{m_{+\,
0}<m_{-\,
0}}\!\!\!\!\!\!\! {\rm d}\Omega \,
a(I_{3\pi}\!=\!2,m_{+\,
0},m_{-\, 0})
\ee
as seen from the table \ref{tab_isospin} so that
\ben
\int\limits_{\rm all}\!\!
{\rm d} \Omega \,
a(I_{3\pi}\!=\!2,m_{+\, 0},m_{-\, 0})=0~.
\label{a2phase}\een

The $\ks$ and $\kl$ decay amplitudes can be obtained using equations
\ref{klks}, \ref{3pi1} and \ref{3pi2}. The
$\ks$ decay amplitude consists of an $I_{3\pi}\!=\!2$ part which is
suppressed by the centrifugal effect, and an $I_{3\pi}\!=\!1$ part
which is suppressed by CP violation, i.e.
\be
A_{\rm S}^{+ \, -\,0}=A_{\rm S} (I_{3\pi}\!=\!2,m_{+\,
0},m_{-\,
0}) + A_{\rm S} (I_{3\pi}\!=\!1,m_{+\, 0},m_{-\,
0})~.
\ee
Using equation \ref{a2phase}, we obtain
\ben
\int\limits_{\rm all}\!\! {\rm d} \Omega \, A_{\rm
S}(I_{3\pi}\!=\!2,m_{+\,
0},m_{-\,
0})=0~.
\label{cancel} \een
The decay amplitude $a(I_{3\pi}\!=\!2)$
can be neglected in the $\kl$ decay amplitude, since it is
doubly suppressed by the centrifugal effect and CP violation:
\be
A_{\rm L}^{+ \, -\,0}= A_{\rm L} (I_{3\pi}\!=\!1,m_{+\,
0},m_{-\, 0}) ~.
\ee
Therefore, both $\ks$ decay amplitudes, i.e. the CP allowed $A_{\rm
S}(I_{3\pi}\!=\!2)$ and the CP violating $A_{\rm S}(I_{3\pi}\!=\!1)$
can interfere with the CP allowed $\kl$ decay amplitude $A_{\rm
L}(I_{3\pi}\!=\!1)$. However, the contribution from the CP allowed
$\ks$ decay amplitude vanishes in the interference once it is
integrated over the entire three-pion phase space as seen from
equation \ref{cancel}. The strong phase shifts $\delta_{3\pi
(I=1)}$ and $\delta_{3 \pi (I=2)}$ are both expected to be small due to
small kinetic energies available for the pions. It follows that the
parameter
\bea
\eta_{+\,-\,0}&=&\frac{\, \int\limits_{\rm all}{\rm d} \Omega \,
 A_{\rm S} (I_{3\pi}\!=\!1,m_{+\, 0},m_{-\, 0})\, A_{\rm
L}^* (I_{3\pi}\!=\!1,m_{+\,
0},m_{-\, 0})\, }
{\int\limits_{\rm all}{\rm d} \Omega \, \left| A_{\rm
L}(I_{3\pi}\!=\!1,m_{+\, 0},m_{-\, 0})\right|^2} \nonumber \\
&\approx& -\varepsilon_{\rm CPT}+\varepsilon_{\rm T} +
\frac{\,i\,}{\,2\,}
\left( 2\,\varphi_{3\pi\left(I=1\right)} + \varphi_{\it
\Gamma}+\overline{\theta}{}_{\rm T}-\theta_{\rm CP}  \right)
\label{eta3pi}\eea
is completely due to CP violation where
$\varphi_{3\pi\left(I=1\right)}$ is the phase of the weak decay
amplitude $a(I_{3\pi}\!=\!1)$ at $m_{+\, 0}\!=\!m_{-\, 0}$.

The contribution from the CP allowed part of the $\ks \rightarrow \pi^+
\pi^- \pi^0$ decay remains in the interference term if the decays are
studied for $m_{+\, 0}>m_{-\, 0}$ and $m_{+\,
0}<m_{-\,0}$ separately. It follows that
\be
\eta_{+\,-\,0}^{m_{+\, 0}>m_{-\, 0}}=
\eta_{+\,-\,0}+\rho
\ee
and
\be
\eta_{+\,-\,0}^{m_{+\, 0}<m_{-\, 0}}=
\eta_{+\,-\,0}- \rho
\ee
where
\ben
\rho = \frac{\int\limits_{m_{+\, 0}>m_{-\,
0}}\!\!\!\!\!\!\! {\rm d} \Omega \,
A_{\rm S}(I_{3\pi}\!=\!2, m_{+\, 0},m_{-\,
0}) A_{\rm L}^* (I_{3\pi}\!=\!1, m_{+\, 0},m_{-\,
0}) }{\int\limits_{m_{+\, 0}>m_{-\, 0}}\!\!\!\!\!\!\!{\rm d} \Omega \,
\left| A_{\rm
L}(I_{3\pi}\!=\!1,m_{+\, 0},m_{-\, 0})\right|^2}
\label{rhodef}\een
{}From various isospin studies of all kaon decays \cite{devlin,jack}, we
expect that
$\rho$ is real and about 20 times larger than $\Re
\left( \eta_{+\,-\,0}\right)$.

The $\ks$-$\kl$ interference term is best studied using the time
dependent rate for initially pure
$\ko$'s decaying into
$\pi^+ \pi^- \pi^0$. When the rate is measured over all the
three-pion phase space, we obtain from equation \ref{initialk0}
\bea
R_{+\,-\,0}(t)&\propto&
e^{-\gl \, t} + \frac{{\it \Gamma}(\ks \rightarrow \pi^+ \pi^- \pi^0)}
{{\it \Gamma}(\kl \rightarrow \pi^+ \pi^- \pi^0)}\, e^{-\, \gs\, t}
\nonumber \\
&&~~~~~~~~~~~~~+ 2\,\left| \eta_{+\,-\,0} \right| \, e^{-\,
\overline{\it
\Gamma}
\, t}
\cos \left( \Delta m \, t + \phi_{+\,-\,0} \right)~.
\nonumber \eea
Note that $\rho$ can be also obtained from $\eta_{+\,-\,0}^{m_{+\,
0}>m_{-\, 0}}$, if the events are restricted only
to $m_{+\, 0}>m_{-\, 0}$,

\subsection{Phase of $\bf \Gamma_{12}$}
The decay matrix element ${\it \Gamma}_{12}$ is given by equation
\ref{g-element}. When unitarity is valid, $\langle {\rm f} | H_{\rm
weak} |i \rangle$ represents the physical decay amplitude. Common
final states between $\ko$ and $\kob$ that should be considered are $2
\pi$ $3 \pi$, $2\pi \gamma$ and $\ell
\pi \nu$ produced by violating the $\Delta Q=\Delta S$ rule. The other
decay modes have negligible branching fractions.  It follows that
\cite{lavula}
\bea
{\it \Gamma}_{12} &\approx&
A_{0}^* \overline{A}{}_{0}+A_{2}^* \overline{A}{}_{2}
+\int d\Omega \left[ A_{2\pi \gamma ({\rm IB})}^* \overline{A}{}_{2\pi
\gamma ({\rm IB})}+A_{2\pi \gamma ({\rm E1})}^* \overline{A}{}_{2\pi
\gamma ({\rm E1})} \right. \nonumber \\
&&~ \left. +A_{2\pi \gamma ({\rm M1})}^*
\overline{A}{}_{2\pi
\gamma ({\rm M1})} \right] + \int d \Omega \left[ A_{3\pi (I=1)}^*
\overline{A}_{3\pi (I=1)} \right. \nonumber \\
&&~~~\left. + A_{3\pi (I=2)}^*
\overline{A}_{3\pi (I=2)} \right] + \int d \Omega \left[ A_{+}^*
\overline{A}{}_+ +
 A_{-}^* \overline{A}{}_-~ \right] , \label{gamma12}
\eea
where IB, E1 and M1 denote the amplitudes for
$\ko$ decaying into $2\pi \gamma$ via inner bremsstrahlung,
direct emission with electric dipole transition and direct
emission with magnetic dipole transition, respectively. The
integrations are done over all necessary phase space including spin.

Compared with the $\ko \rightarrow 3\pi(I=1)$ decay amplitude, the
$\ko \rightarrow 3\pi(I=2)$ decay amplitude can be safely neglected
here as seen in the previous section \cite{devlin,jack}. It can
be shown \cite{hyc} that the direct emission part of the $\ko
\rightarrow \pi \pi \gamma$ decay amplitude is small enough to be
ignored in this discussion. The phase of the inner
bremsstrahlung decay amplitude is almost identical to that of $a_0$.

Equation \ref{gamma12} is now further approximated as
\bea
{\it \Gamma}_{12} &\approx&
\frac{\, 1\,}{\, 2\, }\, {\it \Gamma}_{\rm S} e^{i\, \left( \theta_{\rm
CP} -
\overline{\theta}{}_{\rm T} -2\, \varphi_0 \right)}
\left\{ \rule{0mm}{5mm} 1 + B(\ks \rightarrow 2\pi \gamma)
 \right. \nonumber \\
&&~~~+\left. \frac{\,{\it \Gamma}_{\rm L}\,}{\,{\it
\Gamma}_{\rm S}\,}
\left[ 8 B(\kl \rightarrow \ell^+ \nu \pi^-) \, x
 - B(\kl \rightarrow 3\pi ) e^{i\,2\, \left(\varphi_0 -
\varphi_{3\pi(I=1)} \right)} \right] \right\}
\label{gamma12a} \eea
where $B$'s are the relevant branching ratios and equation
\ref{xparam} is used.

Using equations \ref{gamma12a} and \ref{eta3pi}, the phase of ${\it
\Gamma}_{12}$ is determined to be \footnote{This is identical to
determine the phase of $\varepsilon$ using the Bell-Steinberger
relation \cite{bell-stein}.}
\bea
\varphi_{\it \Gamma}&\approx& -2\, \varphi_0 - \overline{\theta}{}_{\rm
T} +\theta_{\rm CP}
\nonumber \\
&&\!\!\!\! + 2\,\frac{\,\gl \,}{\, \gs \,}\, \left[
4\, B(\kl \rightarrow \ell^+ \pi^- \nu ) \, \Im\left(x\right) -
B(\kl \rightarrow 3\pi)\, \Im\left(\eta_{I=0}- \eta_{3\, \pi} \right)
\right].
\nonumber\eea
Using current experimental values of \cite{ref_pdg}
\be
\Im(x)=-0.003\pm 0.026
\ee
and \cite{3pi1,3pi2,3pi3}
\be
\Im\left(\eta_{3\pi}\right)=0.02 \pm 0.12
\ee
we conclude
\be
\varphi_{\it \Gamma}= -2\, \varphi_0 - \overline{\theta}{}_{\rm
T} +\theta_{\rm CP}\pm 1.8\stimes10^{-4}~.
\ee

Due to this uncertainty, the imaginary part of $\eta_{I=0}$ can
deviate from $\Im \left(\varepsilon _{\rm T}\right)$ as much as
$1.8\stimes 10^{-4}$ even in the absence of CPT
violation as seen from equation \ref{eta0}. Since the real part
of $\varepsilon_{\rm T}$ is $\sim 10^{-3}$, this uncertainty in the
phase of $\eta_{I=0}$ corresponds to $1.6 ^\circ$.

Measurements of $\phi_{+\,-}$ with an uncertainty of much better than
1$^\circ$ are soon to be expected. The standard model prediction is
\be
\varphi_{\it \Gamma}= -2\, \varphi_0 - \overline{\theta}{}_{\rm T}
+\theta_{\rm CP} \pm  {\cal O}\left(<10^{-7}\right),
\ee
thus $\arg\left(\eta_{I=0}\right)=\arg(\varepsilon_{\rm T})
\equiv \phi_{\rm sw}$ if CPT is conserved. However, much better
measurements of $\Im(x)$ and $\Im(\eta_{3\pi})$ are needed in order to
perform a strictly {\bf experimental} test of CPT by comparing the
experimentally measured $\phi_{+\, -}$ and $\phi_{\rm sw}$.

\subsection{Summary}
Let us now summarise important experimental questions on CP
violation in the neutral kaon system discussed in the next chapter:
\begin{itemize}
\item Is the observed CP violation through $\ko$-$\kob$
oscillations accompanied by T violation or CPT violation?\\
This can be studied by measuring non-zero values for $A_{\rm T}$
or $A_{\rm CPT}$ using the semileptonic decays or comparing $\phi_{+\,
-}$ with $\phi_{\rm sw}$. Independent of T or CPT violation, we
expect $\phi_{+\, -}=\phi_{0\, 0}$.

\item Is there CP violation in the decay amplitude?\\
This is the case if we measure a non-zero value for $1-|\eta_{0\,
0}/\eta_{+\, -}|$.

\item Is the phase of ``$\varepsilon$'' obtained from unitarity
identical to $\phi_{\rm sw}=\tan^{-1} \left(2\, \Delta m/ \Delta {\it
\Gamma}\right)$ ? \\
This requires more precise measurements on $\Im (x)$ and $\Im
\left(\eta_{3\pi}\right)$.
\end{itemize}

If CP violation is generated in the ordinary weak interaction, CP
violation in the decay amplitude is expected to appear at some level.
On the other hand, if CP violation occurs only at the higher mass
scale, CP violation can appear only through particle-antiparticle
oscillations.

%%%%%%%%%%%%%%%%%%%%%%%%%%%%%%%%%%%%
%%%%%%%%%%%%%%%%%%%%%%%%%%%%%%%%%%%%
\section{CP Violation Experiments}
\subsection{CPLEAR}
The simplest way to demonstrate CP violation is to compare the
decay of an initially pure $\ko$ into a final
state ``f'' with its CP conjugate process, i.e. an initially pure
$\kob$ decaying into ``$\overline{\rm f}$'' where $\overline{\rm f}$ is
the CP conjugate state of f. CPLEAR is an experiment exactly doing
this.

The $\ko$ and $\kob$ are produced by p$\overline{\rm p}$
annihilation at rest:
\be
{\rm p \overline{p}} \rightarrow \ko {\rm K}^- \pi^+~{\rm or}~
\kob {\rm K}^+ \pi^-~~({\rm branching~ratio:~\sim \!0.2\%~each})~.
\ee
Antiprotons with a flux of $\sim\!10^6$/s and a
momentum of 200~MeV/$c$ extracted from the Low Energy Antiproton Ring
(LEAR) at CERN are stopped in a gaseous hydrogen target with 16~bar
placed in the center of the CPLEAR detector. The target is surrounded
by layers of cylindrical tracking chambers followed by a particle
identification system consisting of a threshold Cherenkov counter
sandwiched by two scintillation counters. The last component of the
detector is a gas sampling electromagnetic calorimeter with lead
converters. All sub-detectors were inserted into of a solenoidal magnet
providing 0.44~T field.

The primary particles, ${\rm K}^{\pm} \pi^{\mp}$ provide the initial
flavour of the neutral kaon and its momentum. The neutral kaon decays
into various final states such as $\pi^+ \pi^-$, $\pi^0 \pi^0$, $\pi^+
\pi^- \pi^0$ and ${\rm e}^{\pm} \pi^{\mp} \nu$.

The experiment is now taking data with its full capability and
the results presented are based on the data taken up to 1992.
Data taking should continue till 1995.

Early results for ${\rm K}\rightarrow \pi^+ \pi^-$ have been
published \cite{ref_cplear1}. Figure
\ref{fig_cplearpi} shows the time dependent CP asymmetry obtained
from recent data. The asymmetry is defined as
\be
A_{+\,-}(t)=\frac{\, \overline{R}{}_{+\,-}(t) - R_{+\,-}(t)\, }
{\, \overline{R}{}_{+\,-}(t) + R_{+\,-}(t)\, }
\ee
where $R(\overline{R})_{+\,-}(t)$ is the time dependent rate for
the initially pure $\ko$($\kob$) decaying into $\pi^+ \pi^-$. A
clear signal of the $\kl$-$\ks$ interference is visible. By fitting the
data with the expected distribution
\be
A_{+\, -}(t)= \Re (\epsilon_{\rm CPT} + \epsilon_{\rm T})
- \frac{\, 2\, \left| \eta_{+\, -}\right|\, e^{\Delta {\it \Gamma}\,
t/2 }
\, \cos\left( \Delta m \, t - \phi_{+\, -} \right) \,}
{1 + \left|\eta_{+\, -} \right|^2 e^{\Delta {\it \Gamma}\, t} }
\ee
which is valid for $t \le 20\, \tau_{\rm S}$ where $\tau_{\rm S}$ is
the $\ks$ lifetime, one obtains \cite{ref_cplear3}
\be
\left.
\begin{array}{rcl}
\left| \eta_{\rm +\, -} \right|&=& \left[ 2.25 \pm
0.07({\rm sta.})\right]
\stimes 10^{-3} \vspace{1mm}\\
\phi_{+ \, -}&=& 44.7^{\circ} \pm 1.3^{\circ} ({\rm
sta.}) \end{array}
\right\}~~{\rm CPLEAR}
\ee
where the final systematic errors are not yet
attributed\footnote{Preliminary systematic errors excluding the effect
due to the uncertainty in
$\Delta m$ are given \cite{ref_cplear3} to be $\pm 0.02
\times 10^{-3}$ for $\eta_{+\, -}$ and
$\pm 0.4^{\circ}$  for $\phi_{+\, -}$.} In the fit, the value of
$\Delta m$ given by the Particle Data Group (PDG) \cite{ref_pdg} was
used.
\begin{figure}
\vspace{76mm}
\hspace{17mm}
\special{illustration apm.eps scaled 450}
\caption{Time dependent CP asymmetry for K$\rightarrow \pi^+ \pi^-$
decays measured by CPLEAR. The solid (dotted) line is the result of a
fit with (without) taking the residual three-body background into
account.}
\label{fig_cplearpi}
\end{figure}

For the  $\pi^+ \pi^- \pi^0$ final state, from the time dependent
asymmetry
\be
A_{+\, -\, 0}(t)= \frac{\, \overline{R}{}_{+\,-\,0}(t) -
R_{+\,-\,0}(t)\, } {\,
\overline{R}{}_{+\,-\,0}(t) + R_{+\,-\,0}(t)\, }
\ee
the result is \cite{ref_cplear3}
\be
\left.
\begin{array}{rcl}
\Re \left(\eta_{+ \, -\, 0}\right)&=&0.002 \pm 0.016({\rm sta.})
\\
\Im \left(\eta_{+ \, -\, 0}\right)&=&0.044 \pm 0.026({\rm sta.})
\end{array}
\right\}~~{\rm CPLEAR}
\ee
where again the systematic errors are under study but claimed to be
much smaller than the statistical errors. Figure
\ref{fig_cplear-etappo} shows $\eta_{+\, -\, 0}$ measured by two
previous experiments \cite{3pi1,3pi3} together with the results from
CPLEAR and E621. The two new
results improve the error by a factor of ten.
\begin{figure}
\vspace{70mm}
\hspace{13mm}
\special{illustration eta+-0.eps scaled 650}
\caption{Measured CP violation parameters for K$\rightarrow \pi^+
\pi^- \pi^0$ decays.}
\label{fig_cplear-etappo}
\end{figure}

CPLEAR measured the CP allowed $\ks \rightarrow \pi^+ \pi^- \pi^0$
decay  amplitude through the interference between the CP allowed $\kl$
and
$\ks$ decay amplitudes. This was done by studying CP asymmetries for
events with
$m_{+\,0}\! >\!m_{-\,0}$ and $m_{+\,0}\!
<\!m_{-\,0}$ separately where $m_{+(-)\,0}$ denotes
the invariant mass between $\pi^+$($\pi^-$) and $\pi^0$. As
explained in the previous chapter, the signs of the interference are
opposite for the two regions.

Figure \ref{fig_cplear-cpok} shows \cite{ref_cplear3} the measured time
dependent CP asymmetry for the two different regions. By fitting the
expected asymmetry
\bea
A_{+\, -\,0}^{m_{+\,0} >(<)m_{-\,0}}(t)&=&
2\, \Re \left( -\varepsilon_{\rm CPT} + \varepsilon_{\rm T} \right)
\nonumber \\
&&\!\!\!\!\!\!\!\!\!\!\!\!\!\!\!\!\!\! -2\, e^{-\Delta {\it \Gamma}\,
t/2}
\,
\left\{ \left[ \Re \left(\eta_{+ \, -\, 0}\right) +\!(-) \rho \right]
\, \cos \Delta m \, t - \Im \left(\eta_{+ \, -\, 0}\right) \,
\sin \Delta m \, t \right\}
\nonumber \eea
a value \cite{ref_cplear3} of $\rho$
\be
\rho = 0.037 \pm 0.011({\rm sta.})~~{\rm CPLEAR}
\ee
was obtained, where $\rho$ given by equation \ref{rhodef} is due to the
interference between CP allowed $\kl$ and
$\ks$ decay amplitudes into the $\pi^+ \pi^- \pi^0$ final state. This
is the first time that the CP allowed $\ks \rightarrow
\pi^+ \pi^-
\pi^0$ decay amplitude has been seen.
\begin{figure}[b]
\vspace{55mm}
%\hspace{5mm}
\special{illustration non-cp(three-pi).eps scaled 630}
\caption{Time dependent CP asymmetries for K$\rightarrow \pi^+ \pi^-
\pi^0$ decays for events with $m_{+\, 0} >m_{-\, 0}$ and $m_{+\, 0}
<m_{-\, 0}$ where $m_{+(-) \,0}$ is the $\pi^+ (\pi^-) \pi^0$
invariant mass of the corresponding decay pions.}
\label{fig_cplear-cpok}
\end{figure}

The study of semileptonic decays provides
different information. Using the time dependent decay rates
for the initially pure $\ko(\kob)$ decaying into ${\rm e}^{\pm
}\pi^{\mp} \nu$, $R(\overline{R})_{\pm}(t)$, the following time
dependent asymmetries are studied:
\be
A_1=\frac{\left[\, \overline{R}{}_- (t) + R_+(t)\right]
-\left[\, \overline{R}{}_+ (t) + R_-(t)\right] }
{\left[\, \overline{R}{}_- (t) + R_+(t)\right]
+\left[\, \overline{R}{}_+ (t) + R_-(t)\right] }
\ee
for $\Delta m$ and $\Re \left( x \right )$,
\be A_2=\frac{\left[\, \overline{R}{}_- (t) +
\overline{R}{}_+(t)\right] -\left[\, R_+ (t) + R_-(t)\right] }
{\left[\, \overline{R}{}_- (t) + \overline{R}{}_+(t)\right]
+\left[\, R_+ (t) + R_-(t)\right] }
\ee
for $\Re \left( -\varepsilon_{\rm CPT}+\varepsilon_{\rm T} \right)$ and
$\Im \left( x \right)$. Preliminary results \cite{ref_cplear3} are
\ben
\begin{array}{rcl}
{\rm CPLEAR}
\\
\Delta m &=&  \left[ 0.524 \pm 0.006({\rm sta.}) \pm 0.002({\rm sys.})
\right]\stimes 10^{10}\,
\hbar {\rm s}^{-1} \\
\Re \left( x \right) &=& -0.024\pm 0.020({\rm sta.}) \pm 0.005({\rm
sys.})
\\
\Im \left( x \right) &=& 0.007\pm 0.008({\rm sta.}) \pm 0.001({\rm
sys.})
\\
\Re \left( -\varepsilon_{\rm CPT}+\varepsilon_{\rm T} \right)&=&
-0.0005 \pm 0.0014({\rm sta.}) \pm 0.0023({\rm sys.})
\end{array}
\label{cpdm}
\een
The current world average of $x$ given by PDG is \cite{ref_pdg}
\be
x=(0.006 \pm 0.018) + i\, (-0.003\pm 0.026)~~{\rm PDG}~.
\ee
The error on $\Im \left( x \right)$ which is of particular interest as
discussed in the previous chapter is improved by a factor of three.
The asymmetry $A_{\rm T}$, a direct signal for CP and T
violation, was also studied and is given by
\be
A_{\rm T}=\frac{\, \overline{R}{}_+(t) - R_- (t)}
{\, \overline{R}{}_+(t) + R_- (t)}=0.000 \pm 0.004({\rm sta.}) \pm
0.008({\rm sys.})~~{\rm CPLEAR}
\ee
assuming a flat distribution of the decay time. If the observed CP
violation is accompanied by T violation, we expect $A_{\rm T}$ to be
$\sim 6.4\! \stimes 10^{-3}$.

\subsection{E621 Experiment}
The experiment E621 is designed to measure CP violation in $\ks
\rightarrow
\pi^+ \pi^- \pi^0$ decays through $\kl$-$\ks$ interference. It is
a fixed target experiment at FNAL using a $\sim \! 60$~m long magnetic
spectrometer with a lead glass calorimeter. The experiment used two
kaon beams of which one was produced very close to
the experiment. The decay time distribution for the
$\pi^+ \pi^- \pi^0$ final states from this beam is given by
\bea
\frac{\, dn_{+\, -\, 0} \,}{\, dt \,}
&=&\frac{N_{\rm L}\, B\left(\kl \rightarrow \pi^+ \pi^- \pi^0
\right)}{\tau_{\rm L}} \left[ e^{-\, \gl \,t}
\right. \nonumber \\
&&~~~+ \left. \left| \eta_{+\, -\, 0}
\right|^2 e^{-\, \gs \, t} + 2\, D \left| \eta_{+\, -\, 0}
\right| e^{-\, \overline{\it \Gamma}\, t} \cos \left( \Delta m \, t +
\phi_{+\, -\, 0} \right)\right]\nonumber
\eea
where $D$ is the fraction of $\ko$ defined as
\be
D=\frac{\, N_{\ko} - N_{\kob} \,}{\, N_{\ko} + N_{\kob} \,}~.
\ee
Since the proton beam produces more $\ko$ than $\kob$, $D\neq
0$ and the $\kl$-$\ks$ interference term remains. The other beam
originated from the experiment delivering only
$\kl$ used to determine the normalization factor. A fit to
the experimental distribution obtained from 18\% of the data taken
gives
\cite{ref_E621}
\be \left. \begin{array}{rcl}
\Re \left( \eta_{+\, -\, 0} \right) &=& 0.005\pm 0.023 \\
\Im \left( \eta_{+\, -\, 0} \right) &=& 0.031\pm 0.030
\end{array} \right\}~~{\rm E621}~.
\ee
The results are shown in figure \ref{fig_cplear-etappo}. By
applying a constraint of $\Re \left(\eta_{+\, -\, 0}\right) =
\Re\left( \eta_{+\, -}\right) = 0.0016$, $\eta_{+\, -\, 0}$ is
obtained to be
\cite{ref_E621}
\be
\Im \left( \eta_{+\, -\, 0} \right) = 0.027 \pm 0.017({\rm sta.})\pm
0.019({\rm sys.})~~{\rm E621}.
\ee

\subsection{Search for Direct CP Violation}
As discussed in the previous chapter, CP violation in the decay
amplitude leads to a nonvanishing $\varepsilon '$. Two experiments,
E731 at FNAL and NA31 at CERN have recently reported their final
analyses on the measurement of $\varepsilon '$. Figure
\ref{fig_detector2} shows both detectors. Note that E731 used a
magnetic spectrometer while the NA31 detector had no magnetic field.
\begin{figure}
\vspace{80mm}
%\special{picture E731.pict}
\special{illustration E731.eps scaled 550}
\vspace{90mm}
%\special{picture NA31.pict}
\special{illustration NA31.eps scaled 650}
\caption{Detector layouts for E731 and NA31 experiments.}
\label{fig_detector2}
\end{figure}

Both experiments measure essentially the ratio $\left|
\eta_{0\,0}/\eta_{+\,-} \right|^2$ which is related to
$\varepsilon '$ as
\ben
\left|\frac{\,\eta_{0\,0}\, }{\, \eta_{+\,-}\, } \right|^2
\approx
 1 - 6\, \Re\left( \frac{\, \varepsilon '\,}{\,
\epsilon \,} \right)
\label{epsp} \een
where equations \ref{reta+-} and \ref{reta00} are used and $\varepsilon
=
\eta_{I=0}$ in our notation. Experimentally, this ratio is given as
\ben
\left|\frac{\,\eta_{0\,0}\, }{\, \eta_{+\,-}\, } \right|^2
=
\left( \frac{\,\epsilon_{0\,0}^{\rm L} n_{0\,0}^{\rm L}\, }{ \,
N_{0\,0}^{\rm L} }
\frac{ \, N_{0\,0}^{\rm S} }
{\,\epsilon_{0\,0}^{\rm S} n_{0\,0}^{\rm S}\, }\right ) \times
\left( \frac{\,\epsilon_{+\,-}^{\rm S} n_{+\,-}^{\rm S}\, }{ \,
N_{+\,-}^{\rm S} }
\frac{ \, N_{+\,-}^{\rm L} } {\,\epsilon_{+\,-}^{\rm L} n_{+\,-}^{\rm
L}\, }\right )
\label{ratio}\een
where $n_{0\,0}^{\rm S(L)}$ and $n_{+\,-}^{\rm S(L)}$ are the numbers
of observed $\ks(\kl) \rightarrow \pi^0 \pi^0$ decays and $\ks(\kl)
\rightarrow \pi^+ \pi^-$ decays respectively, and $N$ denotes the kaon
flux for each detection mode. The experimental correction factors
for each mode are represented by $\epsilon$'s.

NA31 measured $\ks$ and $\kl$ decays separately, but $\pi^+ \pi^-$
and $\pi^0 \pi^0$ final states simultaneously, i.e.
$N_{+\,-}^{\rm L}=N_{0\,0}^{\rm L}$ and
$N_{+\,-}^{\rm S}=N_{0\,0}^{\rm S}$; thus the kaon fluxes
cancel completely in the ratio \ref{ratio}. E731 utilised two
parallel $\kl$ beams where one of them was regenerated to $\ks$. The
regenerator was moved from one beam to the other in between each spill
so that $N_{+\,-}^{\rm S}\propto|\rho|^2 N_{+\,-}^{\rm L}$ and
$N_{+\,-}^{\rm S}\propto |\rho|^2 N_{+\,-}^{\rm L}$ where $\rho$ is the
regeneration amplitude. Then, the kaon fluxes cancel in the ratio
\ref{ratio} too.

The remaining problems are the correction factors. They can be divided
into the following four effects:
\begin{enumerate}
\item geometrical acceptance
\item detector performance; i.e. trigger, energy-momentum resolution,
calibration, detector stability etc.
\item accidentals; overlapping of background events
\item background from other decay modes
\end{enumerate}
\begin{figure}
\vspace{100mm}
\hspace{15mm}
\special{illustration e731_comp.eps scaled 400}
\caption{Comparison between the simulated and real events as a function
of the vertex position $z$ for E731.}
\label{fig_e731-simulation}
\end{figure}
The first effect is given by the beam and detector geometries. The
second correction depends on the beam condition, decay modes and
detection time. The third and fourth corrections are due to the
offline selection where background from the overlapping events or
other decay modes were subtracted. The accidental correction depends
on the beam condition and decay modes and the background correction
depends on the beam condition, decay modes and the detection time.

E731 was designed such that the correction
factors for 2, 3 and 4 cancel in the ratio $|\eta_{0\,0}/\eta_{+\,
-}|^2$. Having two
$\kl$ beams where one of them was regenerated to $\ks$ allowed to
take $\kl$ and $\ks$ decays simultaneously in the same experimental
environment. This makes correction factors to be $\epsilon_{+\,-}^{\rm
S} \approx \epsilon_{+\,-}^{\rm L}$ and
$\epsilon_{0\,0}^{\rm S} \approx \epsilon_{0\,0}^{\rm L}$ {\bf except
for the geometrical acceptance}, hence they cancel in the ratio
\ref{ratio}. Due to the difference in the lifetimes, $\ks$ and $\kl$
have different decay vertex distributions. The accepted decay regions
for the $\pi^+ \pi^-$ and $\pi^0 \pi^0$ final states in the
E731 detector are not identical. Therefore, the detector acceptances
are completely different for all four modes and the geometrical
correction factors do not cancel in the ratio \ref{ratio}.

The effect due to the geometrical acceptance was studied by
simulation. A very detailed simulation programme was
tuned using $\kl \rightarrow \pi^{\pm} {\rm e}^{\mp} \nu$ events which
were also accepted by the $\pi^+ \pi^-$ trigger. The $\kl \rightarrow
3\, \pi^0$ events were used for tuning the neutral decay simulation.
Figure \ref{fig_e731-simulation} \cite{ref_E7311,ref_E7312} shows the
comparison between the simulated and measured kaon decay vertex
distribution for the
$\pi^{\pm} {\rm e}^{\mp} \nu$ events. The tuned
simulation programme was used to correct the acceptance for $\pi^+
\pi^-$ and
$\pi^0 \pi^0$ events.
\begin{figure}
\vspace{70mm}
\hspace{4mm}
\special{illustration na31_sp.eps scaled 700}
\caption{Energy spectra and weighted decay vertex distributions for
$\ks$ and
$\kl$ beams observed by NA31.}
\label{fig_na31}
\end{figure}

NA31 optimised their detector to cancel the effect from the
geometrical acceptance. The $\ks$ beam was produced by a moving
target which simulated the decay point distribution of the $\kl$ beam.
The incident proton momenta were chosen such that the $\kl$ and $\ks$
beams had similar momentum spectra. Thus, the geometrical acceptances
are almost identical between $\kl$ and $\ks\rightarrow \pi^+ \pi^-$
and between $\kl$ and $\ks\rightarrow \pi^0  \pi^0$.
Figure \ref{fig_na31} \cite{ref_NA312} shows the energy
spectra and the weighted vertex distributions for the $\kl$ and $\ks$
beams. All the other effects in the correction factors, in principle,
do not cancel in the ratio
$|\eta_{0\,0}/\eta_{+\, -}|^2$ since
$\kl$ and
$\ks$ decays were taken at different time with different conditions.
The intensities of the primary protons were adjusted to achieve
similar background conditions between
$\ks$ and $\kl$ beams in order to minimise the difference.
\begin{table}
\begin{center}
\caption{Event statistics for NA31 and E731 after subtracting all
the background.}
\label{tab_data}
\begin{tabular}{lrrr}
\multicolumn{1}{c}{Decay Mode} & \multicolumn{2}{c}{NA31} &
\multicolumn{1}{c}{E731}\\
 &\multicolumn{1}{c}{1986}& 1988+1989 & \\ \hline
$\kl \rightarrow \pi^0 \pi^0 \rule{0mm}{4mm}$ &$109\stimes 10^3$
&$319\stimes 10^3~$ &$410\stimes 10^3$ \\
$\kl \rightarrow \pi^+ \pi^-$ &$295\stimes 10^3$ &$847\stimes 10^3~$
&$327\stimes 10^3$ \\
$\ks \rightarrow \pi^0 \pi^0$ &$932\stimes 10^3$ &$1322\stimes
10^3~$ &$800\stimes 10^3$ \\
$\ks \rightarrow \pi^+ \pi^-$ &$2300\stimes 10^3$ &$3241\stimes
10^3~$ &$1061\stimes 10^3$
\end{tabular}
\end{center}
\end{table}

Table \ref{tab_data} summarises the statistics of the two
experiments after background subtraction
\cite{ref_E7311,ref_NA312,ref_NA311}. NA31 took data in 1986,
1988 and 1989. The result from the 1986 data has been published
\cite{ref_NA311}. Some detector improvements were made for the 1988
and 1989 data taking. E731 took all the data during the run from 1987
to 1988. Results obtained from part of the data has been published
previously \cite{ref_E731old}. Statistical limitations come from the
$\kl$ decays in both experiments (for E731, it was the
$\pi^+ \pi^-$ mode and for NA31 it was the $\pi^0 \pi^0$ decay mode).
\begin{table}[h]
\begin{center}
\caption{Subtracted background.}
\label{tab_background}
\begin{tabular}{lrrr}
\multicolumn{1}{c}{Decay Modes} & \multicolumn{2}{c}{NA31} &
\multicolumn{1}{c}{E731}\\
 &\multicolumn{1}{c}{1986}& 1988+1989 & \\ \hline
\multicolumn{4}{c}{three-body decay background} \\
$\kl \rightarrow \pi^0 \pi^0$ &$4.0\%$ & $2.67 \%~~~$ & $1.78 \%$ \\
$\kl \rightarrow \pi^+ \pi^-$ &$0.6\%$ & $0.63 \%~~~$ & $0.32 \%$ \\
$\ks \rightarrow \pi^0 \pi^0$ &$<0.1\%$ & $0.07\%~~~$ & $0.049\%$\\
\hline
\multicolumn{4}{c}{effect due to the regenerator} \\
$\kl \rightarrow \pi^0 \pi^0$ &\multicolumn{1}{c}{-}
&\multicolumn{1}{c}{-} & $2.26 \%$\\
$\ks \rightarrow \pi^0 \pi^0$ &\multicolumn{1}{c}{-}
&\multicolumn{1}{c}{-} & $2.53 \%$\\
$\ks \rightarrow \pi^+ \pi^-$ &\multicolumn{1}{c}{-}
&\multicolumn{1}{c}{-} & $0.155
\%$\\ \hline
\multicolumn{4}{c}{effect due to the trigger plane} \\
$\kl \rightarrow \pi^0 \pi^0$ &\multicolumn{1}{c}{-}
&\multicolumn{1}{c}{-} & $1.12 \%$\\
$\ks \rightarrow \pi^0 \pi^0$ &\multicolumn{1}{c}{-}
&\multicolumn{1}{c}{-} & $0.264 \%$
\end{tabular}
\end{center}
\end{table}

Table \ref{tab_background} summarises the subtracted
background \cite{ref_E7311,ref_NA312,ref_NA311}. In
addition to the usual three body decays of kaons, E731 studied
additional sources of background. One is due to the incoherent
regeneration of the $\kl$ beam in the regenerator. This has to be
subtracted in the $\ks$ decays. For the $\pi^0 \pi^0$ decay mode, the
incoherent regeneration introduces a background in the $\kl$ decays as
well. Due to the worse vertex resolution for the $\pi^0 \pi^0$ decays,
some of the reconstructed $\ks$ decay vertices fall on the $\kl$ beam
spot taken as $\kl$ decays. Another source of background was the
trigger plane for the charged decay mode which affected only the
$\pi^0 \pi^0$ decays (see figure \ref{fig_detector2}).
\begin{table}
\begin{center}
\caption{List of systematic uncertainties in $|\eta_{0\,
0}/\eta_{+\, -}|^2$.}
\label{tab_systematics}
\begin{tabular}{lrrr}
 & \multicolumn{2}{c}{NA31} &
\multicolumn{1}{c}{E731}\\
 &\multicolumn{1}{c}{1986}& 1988+1989 & \\ \hline
energy calibration \rule{0mm}{4mm}&$3.0 \stimes 10^{-3}$&$1.3\stimes
10^{-3}$&$0.96 \stimes 10^{-3}$\\
accidental correction &$2.0 \stimes 10^{-3}$&$1.4 \stimes 10^{-3}$
&$0.64 \stimes 10^{-3}$\\
acceptance & $1.7 \stimes 10^{-3}$&$1.0 \stimes 10^{-3}$&
$0.71 \stimes 10^{-3}$ \\
background & & & \\
$~~~\kl \rightarrow \pi^0 \pi^0$ &
$2.0 \stimes 10^{-3}$&$1.3 \stimes 10^{-3}$ &$0.36 \stimes 10^{-3}$ \\
$~~~\kl \rightarrow \pi^+ \pi^-$ &
$2.0 \stimes 10^{-3}$&$1.0 \stimes 10^{-3}$ &$0.17 \stimes 10^{-3}$ \\
trigger and anti $\ks$ counter & & &\\
${\rm ~~~inefficiencies}$ & $1.0 \stimes 10^{-3}$
&$ 0.9 \stimes 10^{-3}$ &\multicolumn{1}{c}{-} \\
wire chamber inefficiencies &\multicolumn{1}{c}{-} & $1.0 \stimes
10^{-3}$ &\multicolumn{1}{c}{-} \\
regenerator related corrections &\multicolumn{1}{c}{-}
& \multicolumn{1}{c}{-}& $0.59 \stimes 10^{-3}$ \\
trigger plane efficiency & \multicolumn{1}{c}{-} &
\multicolumn{1}{c}{-}&
$0.73 \stimes 10^{-3}$ \\
beam scattering &\multicolumn{1}{c}{-}&\multicolumn{1}{c}{-}& $0.18
\stimes 10^{-3}$
\\ \hline total \rule{0mm}{4mm}& $ 5.0 \stimes 10^{-3}$& $3.0 \stimes
10^{-3}$ &$ 1.71 \stimes 10^{-3}$
\end{tabular}
\end{center}
\end{table}

Table \ref{tab_systematics} lists systematic
uncertainties in the ratio $\left| \eta_{0\,0} /\eta_{+\, -}
\right|^2$ for the two experiments
\cite{ref_E7311,ref_NA312,ref_NA311}. Figure \ref{fig_total} shows
$|\eta_{0\,0} /\eta_{+\, -}|$ for E731(final) and NA31
results with 1986 data and with 1987+1988 data
\cite{ref_E7311,ref_NA312,ref_NA311}. The statistic and systematic
errors are separately drawn indicating that NA31 is limited
by systematics and E731 is statistically limited.

The parameter $\Re( \varepsilon ' / \varepsilon)$ is determined
from equation \ref{epsp}. For NA31, the common systematic
errors in the 1986 and 1988+1989 data have been taken into
account when the two results were combined \cite{ref_NA312}. It follows
that
\be
\Re \left( \frac{\, \varepsilon' \,}{\varepsilon} \right) =
\left\{ \begin{array}{ll}
(23.0 \pm 6.5) \stimes 10^{-4}& \mbox{\rm NA31(1986+1988+1989)}
\\
(7.4 \pm 5.9) \stimes 10^{-4}& \mbox{\rm E731}
\end{array}\right.
\ee

Taken Individually, the two experiments may lead to two different
conclusions. The NA31 result can be interpreted that
$\Re(\varepsilon'/\varepsilon)>0$ with a significance of more than
three standard deviations, a strong indication of CP violation in the
decay amplitude. The result from E731 is compatible with
$\Re(\varepsilon'/\varepsilon)=0$, thus with no direct CP violation.
However, the probability for the two results being statistically
compatible is 7.6\% which is not negligible. If we combine the two
results and scale the combined error in the way done by the Particle
Data Group \cite{ref_pdg}, we obtain
\be
\Re \left( \frac{ \varepsilon'}{\varepsilon} \right) = (14.4 \pm
7.8) \stimes 10^{-4}~~{\rm E731+NA31} \nonumber
\ee
which gives a significance of less than two standard deviations for a
non-zero value for $\Re(\varepsilon'/\varepsilon)$.

We conclude that
CP violation in the decay amplitude is not yet established from the
measured $\Re(\varepsilon'/\varepsilon)$. Experiments with a better
sensitivity are required in order to resolve the discrepancy between
NA31 and E731 which deviate by 1.8 standard deviations.
\begin{figure}
\vspace{73mm}
\hspace{20mm}
\special{illustration eta/eta.eps scaled 555}
\caption{The measured ratio $\left| \eta_{0\,0}/\eta_{+\, -} \right|$
by NA31 and E731.}
\label{fig_total}
\end{figure}

\subsection{CPT Test by E731}
As described in the previous chapter, the phases of $\eta_{+\, -}$ and
$\eta_{0\, 0}$ are expected to be almost identical to
\be
\phi_{\rm sw}=\tan^{-1} \left( \frac{\, 2\, \Delta m\,}{\Delta {\it
\Gamma} }\right)  =43.7^{\circ} \pm 0.2^{\circ}~,
\ee
if CPT is conserved in $\ko$-$\kob$ oscillations. We used the
current average value\footnote{The error on $\Delta m$ is scaled by
1.24 applying the PDG recipe \cite{ref_pdg}.} of $\Delta m$
\cite{ref_dm1,ref_dm2,ref_dm3}
\ben
\Delta m =(0.5352\pm0.0031) \stimes 10^{10} \, \hbar \, {\rm
s}^{-1}~
\label{dm-world}
\een
and the PDG value \cite{ref_pdg}
\ben
\tau_{\rm S}=(0.8922 \pm 0.0020) \stimes 10^{-10} \, {\rm s}~~{\rm
PDG}
\label{ts-pdg}
\een
for obtaining $\Delta {\it \Gamma}$.

The current world average value for
$\phi_{+\, -}$ given by PDG \cite{ref_pdg} is
\be
\phi_{+\, -}= 46.5^{\circ} \pm 1.2^{\circ}~~{\rm PDG},
\ee
which is 2.3 standard deviations apart from $\phi_{\rm sw}$.

Using the same data sample as in the $\varepsilon '$ analysis, E731
asked whether the data were consistent with CPT conservation
\cite{ref_E7313}. This requires to measure $\Delta m$, $\gs$, $\phi_{+
\, -}$ and the phase difference $\phi_{0\, 0}-\phi_{+\, -}$. Since $\gs
\gg
\gl$, the value of $\gl$ is much less important. The following
describes how the data were analysed.

After correcting for the detector acceptance, the decay rates of the
$\kl$ beam without regenerator and that behind the regenerator with a
given momentum $p$ and at a decay position $z$ are given by
\ben
\frac{\, d^2  n_{\rm L}\, }{\,dz\, dp\,}
= F(p) \left| \eta_{\rm f} \right|^2 e^{ -\, \gl \,
z /
\beta
\,
\gamma \, c}
\label{E731phase0}\een
and
\bea
\frac{\, d^2  n_{\rm reg}\, }{\,dz\, dp\,}
&=&  t\, F(p) \left [ \left| \rho(p) \right|^2 e^{ -\,
\gs \, z / \beta
\,
\gamma
\, c} +\left| \eta_{\rm f} \right|^2 e^{ -\, \gl \, z / \beta \, \gamma
\, c} \right.
\nonumber \\
&&\left. ~+2\, \left| \eta_{\rm f} \rho(p) \right|
e^{-\, \overline{\it \Gamma}\, z / \beta \, \gamma \, c}
\cos\left( \Delta m\, z / \beta \, \gamma \, c - \phi_{\rm f} +
\phi_{\rho}\right) \right] \label{E731phase}
\eea
respectively, where
the final state ``f'' represents both
$\pi^+
\pi^-$ and $\pi^0 \pi^0$ and $\eta_{\rm f}$ is the CP violation
parameter. The parameter $t$ is the relative transmission of the kaon
to the $\kl$ beam without regenerator, which was constrained to the
value measured by $\pi^+ \pi^- \pi^0$ and $3 \pi^0$ final states.
$F(p)$ is the kaon flux, $\beta\, \gamma$ is the Lorentz boost factor
and $\rho(p)$ and $\phi_{\rho}$ are the regeneration amplitude and its
phase respectively.

Apart from the trivial factor coming from the propagation of the
kaon in the regenerator, the modulus of the regeneration amplitude is
assumed to be given by a power of the
$\ks$ momentum as
\be
\left| \rho(p)\right| \propto b \left(\frac{p}{70 {\rm GeV}/c}\right)^a
\ee
where $a$ and $b$ are free parameters\footnote{Detailed discussions
on this subject are found in \cite{ref_E7312} and \cite{ref_E7313}}.
This assumption is based on the observation that a single Regge
trajectory ($\omega$) is dominant in the regeneration process at high
energies. The validity of this power law with a single exponent was
tested with the data and quoted errors include a possible deviation.
Then, the phase of the regeneration amplitude is given by the exponent
$a$ as
\be
\phi_{\rho}= \frac{\, a-1 \,}{2} \, \pi + \phi_{\rm G}
\ee
due to analyticity where $\phi_{\rm G}$ is coming from the propagation
of the kaon in the regenerator. With this relation, the regeneration
phase is determined from the term proportional to $|\rho|^2$ in
equation
\ref{E731phase}. E731 tested the validity of the analyticity relation
experimentally and found it to be better than the other experimental
uncertainties.

The experimentally observed decays were binned in various $z$ and $p$
values and equations \ref{E731phase0} and \ref{E731phase} were used to
fit the data. First, the phases of the CP violation parameters are
fixed to
\be
\phi_{+\,-}=\phi_{0\,0}=\tan^{-1}\left(\frac{2\, \Delta m}{\Delta
{\it \Gamma} }\right)
\ee
and $\pi^+ \pi^-$ and $\pi^0 \pi^0$ decay modes are fitted separately
with $a$, $b$, $\gs$ and $\Delta m$ as free parameters with the
$\eta$'s constrained to their known values. This is the first time
that $\Delta m$ and $\tau_{\rm S}$ are measured using the
$\pi^0 \pi^0$ decay mode with a comparable precision as the results
obtained from the $\pi^+ \pi^-$ decay mode. Combining the  results from
the $\pi^+ \pi^-$ and
$\pi^0 \pi^0$ decay modes, it follows that \cite{ref_E7313}
\ben \left.
\begin{array}{rcl}
\tau_{\rm S}&=&\gs^{-1}=(0.8929 \pm 0.0016) \stimes 10^{-10}\, {\rm s}
\vspace{1mm}\\
\Delta m &=&(0.5286\pm0.0028) \stimes 10^{10} \, \hbar \,{\rm
s}^{-1}
\end{array}
\right\}{\rm E731}~.
\label{E731deltam}
\een

The E731 result on the $\ks$ lifetime is in perfect agreement with
the PDG value given in equation \ref{ts-pdg}. The central value of
$\Delta m$ obtained by E731 is 2.2 standard deviations smaller than the
world average in equation \ref{dm-world}, although the two values are
only 1.6 standard deviations apart.

The phase difference $\phi_{0\, 0} - \phi_{+ \, -}$ is
measured by fitting both $\pi^+ \pi^-$ and $\pi^0 \pi^0$ decay modes
simultaneously while fixing $\gs$ and $\Delta m$ to their measured
values and with $a$, $b$, $\phi_{+\, -}$, $\phi_{0\, 0} - \phi_{+
\, -}$ and
$\Re(\varepsilon' /\varepsilon)$ as free parameters. The result is
\cite{ref_E7313}
\be
\phi_{0\, 0} - \phi_{+\, -} = -1.6^{\circ} \pm
1.2^{\circ}~~{\rm E731}
\ee
which provides a better limit on the phase difference than from the
current PDG average \cite{ref_pdg}
\be
\phi_{0\, 0} - \phi_{+\, -} = -0.1^{\circ} \pm 2.0^{\circ}~~{\rm
PDG}.
\ee
\begin{figure}[bt]
\vspace{71mm}
\hspace{14mm}
\special{illustration dm-phi.eps scaled 650}
\caption{The $\Delta m$ dependence of the extracted phase $\phi_{+\,
-}$ from the observed $\kl$-$\ks$ interference for three different
experiments.}
\label{fig_phidm}
\end{figure}

Lastly, $\Delta m$ was treated also as a free parameter in the fit to
extract $\phi_{+\,-}$ and the fit gives \cite{ref_E7313}
\be
\phi_{+\, -}=42.2^{\circ} \pm 1.4^{\circ}~~{\rm E731}.
\ee
At the same time, the fit gives $\Delta m =(0.5257 \pm 0.0049) \stimes
10^{10}
\, \hbar \,{\rm s}^{-1}$ \cite{ref_E7312} which is in good agreement
with their own value shown in \ref{E731deltam}. With the E731 values
for
$\gs$ and
$\Delta m$,
$\phi_{\rm sw}$ is determined to be
\be
\phi_{\rm sw}=43.4^{\circ} \pm 0.2^{\circ}~~{\rm E731}~.
\ee

In summary, E731 concludes that their data are indeed compatible with
CPT conservation and
\be
\phi_{+ \, -} \approx \phi_{0 \, 0} \approx \phi_{\rm sw}~~{\rm E731}.
\ee

How can we accommodate this new results with previous measurements?
The E731 data clearly prefers a lower value of $\Delta m$ as
shown from the third analysis. The value of $\phi_{+\,-}$ is extracted
from the $\kl$-$\ks$ interference term proportional to
$\cos \left[ \Delta m \, t - \phi_{+\, -} (+ \phi_{\rho}) \right]$.
If there were equal numbers of events at all decay times, $\Delta m$
and $\phi_{+\, -}$ could be decoupled by measuring more than half
of the oscillation length since $\Delta m$ gives the frequency and the
$\phi_{+\,-}$ gives the phase of the oscillation. However, more weight
is given to earlier decay times due to the exponential decay law
introducing a positive correlation between the measured $\Delta m$ and
$\phi_{+\, -}$. Figure \ref{fig_phidm} shows the extracted $\phi_{+\,
-}$ as a function of $\Delta m$ for two experiments
\cite{ref_gew,ref_E731p}. The differences in the slopes are due to
different methods, one using vacuum regeneration and the other
using a regenerator. Also shown is the superweak phase.

Curiously, the two experimental results and the superweak phase seem to
agree at
$\Delta m$ which is somewhat smaller than the current world average.
This conclusion remains valid even including more results of
$\phi_{+\, -}$ \cite{ref_phio1,ref_phina31}, which are less accurate.

Figure \ref{fig_dm} shows three previous $\Delta m$ measurements which
are used to obtain the world average \cite{ref_dm1,ref_dm2,ref_dm3}.
Two experiments, C.~Geweniger et al. and S.~Sjesdal et al. used the
same detector with an additional common systematic error of 0.0015.
The third experiment, M.~Cullen et al. gives a higher value than the
other two. The E731 result is also shown.

Whether the true value of $\Delta m$ is indeed about two standard
deviations lower than the current value needs an independent
confirmation from other experiments, in particular those without
regenerator. CPLEAR measuring simultaneously $\Delta m$ using
semileptonic decays and $\phi_{+ \, -}$ from the $\pi^+ \pi^-$ decay
modes will make an important contribution. It is interesting to note
that its preliminary value of
$\Delta m$ given in equations \ref{cpdm} is also smaller than the
world average.
\begin{figure}
\vspace{69 mm}
\hspace{14mm}
\special{illustration Delta-m.eps scaled 650}
\caption{The $\kl$-$\ks$ mass difference measured by various
experiments used by the Particle Data Group to obtain the world
average and the E731 result.}
\label{fig_dm}
\end{figure}

The possibility that a wrong value of $\gs$ is responsible for this
problem is not likely. Firstly, E731 gives a value of $\gs$ which is
perfectly consistent with the world average. Secondly, a different
value of $\gs$ does not give a consistent answer. In order to
demonstrate this, let us consider the two experiments which give
explicit
$\Delta m$ and $\gs$ dependences of their extracted $\phi_{+\,-}$
values \cite{ref_E731p,ref_phina31}
\be
\phi_{+\, -} \! =\! \! \left\{ \!\!\!
\begin{array}{lr}
 46.9^{\circ} \pm 1.6^{\circ}\! + \! \left( \frac{\displaystyle
\rule{0mm}{3mm}
\tau_{\rm S}\,}{\displaystyle \rule{0mm}{3mm} 0.8922} -1 \right)
\!\! \times \!
270^{\circ}\! + \! \left( \frac{\displaystyle \rule{0mm}{3mm} \Delta m
\,}{\displaystyle \rule{0mm}{3mm} 0.5351}-1\right) \!\! \times \!
310^{\circ} & {\rm NA31}
\vspace{1 mm} \\
42.21^{\circ} \pm 0.9^{\circ}\! - \! \left(  \frac{\displaystyle
\rule{0mm}{3mm}
\tau_{\rm S}\,}{\displaystyle \rule{0mm}{3mm} 0.8922}  -1 \right)
\!\! \times \! 410^{\circ}\! +\! \left(
\frac{\displaystyle \rule{0mm}{3mm} \Delta m\,}{\displaystyle
\rule{0mm}{3mm} 0.5286}-1
\right)
\!\! \times \! 100^{\circ} & {\rm E731}
\end{array} \right.\ee
and as the third condition, we take the CPT invariance, i.e.
\be
\phi_{+\, -}=\tan^{-1}\left(\frac{\, 2\, \Delta m \,}{\, \Delta {\it
\Gamma}\, }\right)~.
\ee
The three equations give a solution
\bea
\tau_{\rm S}&=&0.8900 \stimes 10^{-10} \, {\rm s} \nonumber \\
\Delta m &=& 0.5303 \stimes 10^{10} \, \hbar \,{\rm s}^{-1} \nonumber
\\
\phi_{+\, -}&=&\phi_{\rm sw}=43.4 ^{\circ} \nonumber
\eea
which could be an ``educated guess''.

\subsection{Summary}
CP violation is still observed only in the neutral kaon system, with
four different processes:
\begin{enumerate}
\item Charge asymmetry in the semileptonic $\kl$ decays.
\item CP violating $\kl \rightarrow \pi^+ \pi^-$ decay amplitude.
\item CP violating $\kl \rightarrow \pi^0 \pi^0$ decay amplitude.
\item CP violating $\kl \rightarrow \pi^+ \pi^- \gamma$ decay
amplitude recently observed by E731\cite{ref_E7314}.
\end{enumerate}
CP violation in the oscillation must be accompanied with either T
violation or CPT violation or both. The preliminary measurement of
$A_{\rm T}$ with $0.000 \pm 0.009$ (sensitive only to CP+T violation)
by CPLEAR does not yet show a direct signal of T violation with the
present statistics. However, it should be seen with the expected
increase in statistics in the near future.

In order to test CPT indirectly from the phase of $\eta_{+\, -}$,
the phase of ``$\varepsilon$'' must be calculated using the unitarity
relation. With new values for $\Im(x)$ measured by CPLEAR and
$\eta_{+\, -\, 0}$ measured by E621 and CPLEAR,
the phase of ``$\varepsilon$'' is equal to $\phi_{\rm sw}$ within
$\sim 0.4^\circ$.  Further improvements on $x$ and $\eta_{+\, -\, 0}$
are still needed for future CPT test.

New measurements on $\tau_{\rm S}$, $\Delta m$, $\phi_{+\, -}$ and
$\phi_{0\, 0}-\phi_{+\, -}$ by E731 are in perfect agreement
with CPT conservation. The discrepancy between $\phi_{+\, -}$ and
$\phi_{\rm sw}$ with previous measurements at a level of
2.3 standard deviations can be explained by a new
value of
$\Delta m$ which would be about one standard deviation lower than the
current average value given by the Particle Data Group. Future
measurements will answer this question.

NA31 measured a positive value of $\Re( \varepsilon '
/ \varepsilon)=2.3\times 10^{-3}$ with a significance of more than
three standard deviations. E731 gives $\Re( \varepsilon '
/\varepsilon)=7.4\times 10^{-4}$ with a one standard
deviation significance. By combining the two results, we conclude
that a non-zero value of $\Re(\varepsilon ' /\varepsilon )$ is not
yet established, i.e. no sign of direct CP violation. Most
recent theoretical calculations for $\Re( \varepsilon '
/\varepsilon)$ \cite{buras} tend to favour the value obtained by E731.
However, uncertainties in the hadronic matrix elements involving the
penguin diagram are still too large to be reliable.

%%%%%%%%%%%%%%%%%%%%%%%%%%%%%%%%%%%%%%%%%%%%%%%%%%%%%%
%%%%%%%%%%%%%%%%%%%%%%%%%%%%%%%%%%%%%%%%%%%%%%%%%%%%%%%
\section{Future Prospects}
\subsection{Neutral Kaon System}
Experimentally, the kaon system will continue to dominate the
field of CP violation for a while. CPLEAR plans to
improve the statistical errors by roughly a factor of four by 1995
\cite{ref_cplear3}. This will for example allow to measure
$\phi_{+\, -}$ to better than $0.5^{\circ}$ and $\Delta m$ within
$0.002 \stimes 10^{10}\, \hbar \, {\rm s}^{-1}$. These two
measurements will improve the indirect CPT test in the $\ko$-$\kob$
oscillation. An expected error of $A_{\rm T}$ of $0.001$ will clearly
establish T violation in the $\ko$-$\kob$ oscillation. Although CPLEAR
will not reach the sensitivity to observe CP violation in the
K$\rightarrow \pi^+ \pi^- \pi^0$ decay, a better limit on $\eta_{+ \,
-\, 0}$ together with an improved limit on $\Im \left( x\right)$ will
lower the uncertainty in the phase value of ``$\varepsilon$'' using the
unitarity relation. Without this, measuring $\phi_{+\, -}$ with a
precision better than 0.5$^\circ$ does not really provide a
rigorous test of CPT.

E773 is a dedicated experiment at FNAL to measure the phases of
the $\eta$'s with the E731 detector. Results are expected on $\phi_{+
\, -}$ and $\phi_{0\, 0} -\phi_{+\, -}$ very soon with errors smaller
than
$1^{\circ}$ and $0.5^{\circ}$ respectively \cite{Gollin}.

In the field of the three-pion decay mode, E621 will present a
better limit on $\eta_{+\, - \, 0}$ with full statistics. E621 should
be able to measure the CP allowed $\ks \rightarrow \pi^+ \pi^- \pi^0$
decay amplitude as well.

Next generation $\Re \left( \varepsilon ' / \varepsilon
\right)$ experiments are already under construction: NA48
\cite{NA48} at CERN and E832(KTEV) at FNAL \cite{KTEV}. Both
experiments use magnetic spectrometers with very good electromagnetic
calorimeters: CsI crystals for KTEV and liquid Krypton for NA48. Both
have $\ks$-$\kl$ double beams: Two different targets with tag
counters for NA48 and two
$\kl$ beams with a much improved regenerator for KTEV. Both will
be able to take all four decay modes $\kl,\ks \rightarrow \pi^+ \pi^-,
\pi^0 \pi^0$ at the same time reducing systematic uncertainties
considerably. Both experiments expect to start data taking in 1995 and
the goal is to achieve a sensitivity of less that $10^{-4}$.

Various rare kaon decay experiments being continued at BNL and KEK are
to some extent also sensitive to CP violation in other decay channels.
Further discussion can be found in the contribution from J.~Ritchie in
these proceedings.

A new $\phi$ factory operating at $\sqrt{s}\approx 1 {\rm
GeV}$ under construction in Frascati (DA$\Phi$NE) \cite{franz} with
a designed luminosity of $\sim 10^{32}\, {\rm cm}^{-2}\, {\rm s}^{-1}$
will provide $5 \stimes 10^9$ tagged $\ks$ and $\kl$ (also $\sim
2\stimes 10^{10}$ tagged K$^{\pm}$) in one year. This is well suited
to make global studies of CP, T and CPT violation, extending CPLEAR
results \cite{sanda2}. A unique feature of the
$\phi$ decays producing pure
$\ks$-$\kl$ (or
$\ko$-$\kob$) initial states\footnote{A similar situation occurs in
the $\Upsilon$(4S) decays into B$\overline{\rm B}$.} will allow
to test quantum mechanics. This will be only the place able to
study rare
$\ks$ decays, e.g. CP violating $\ks \rightarrow 3\pi^0$ decays
($\sim 30$ decays in one year). A sensitivity of $\sim 10^{-4}$ for
measuring the real and imaginary part of $
\varepsilon' / \varepsilon$ with the KLOE detector seems possible. More
details and further references can be found in \cite{franz}.

The Main Injector ring at FNAL can provide very intensive kaon
beams which may allow to construct a detector capable of measuring
$\Re \left( \varepsilon ' / \varepsilon \right)$ with a
sensitivity much better than $10^{-4}$ \cite{KAMI}.

\subsection{Charged Kaons}
In the standard model, a CP violation effect larger than $\varepsilon
'$ is expected in the charged kaon decay rates asymmetry
\be
\frac{{\it \Gamma}_{{\rm K}^+ \rightarrow \pi^+ \pi^+ \pi^-}
- {\it \Gamma}_{{\rm K}^- \rightarrow \pi^+ \pi^- \pi^-}}
{{\it \Gamma}_{{\rm K}^+ \rightarrow \pi^+ \pi^+ \pi^-} + {\it
\Gamma}_{{\rm K}^- \rightarrow \pi^+ \pi^- \pi^-}}~.
\ee
The slope parameter in the
Dalitz plot is also expected to be different in this decay mode between
K$^+$ and K$^-$. With DA$\Phi$NE or at BNL \cite{zeller}, such an
effect could be investigated.

\subsection{The $\bf \Lambda$ System}
A CP asymmetry in the decay asymmetries of $\alpha$ for $\Lambda
\rightarrow {\rm p} \pi^-$ and $\overline{\alpha}$
for $\overline{\Lambda} \rightarrow \overline{\rm p} \pi^+$ given by
\be
A_{\alpha}=\frac{\, \alpha + \overline{\alpha} \, }
{\, \alpha - \overline{\alpha} \, }
\ee
is non-zero if there exists an interference between the proton-pion s-
and p-wave amplitudes. Such an interference could be possible
in the standard model where the penguin diagrams contribute
differently to the s- and p-waves. This effect is expected to be less
than $10^{-4}$ \cite{donog}. The experiment PS185 at CERN measured
\cite{P185}
\be
A_{\alpha}=-0.07 \pm 0.09
\ee
using $\Lambda$-$\overline{ \Lambda}$ pairs produced by
p$\overline{\rm p}$ annihilations in flight. A proposal submitted
to FNAL \cite{FLAM} to measure this asymmetry using $\Lambda$ and
$\overline{\Lambda}$ from hyperon decays hopes for a sensitivity of
$10^{-4}$. Possible new facilities such as Super LEAR or a Tau-charm
factory can provide a clean and intense source of
$\Lambda$-$\overline{\Lambda}$.

\section{B-Meson System}
The Physics of B-mesons are discussed by A.~Sanda in
these proceedings.
\subsection{Formalism}
We just present a short description of the neutral B-meson system in
order to make a comparison to the neutral kaon system. For
simplicity, we assume that CPT is conserved \cite{sanda}. The two
mass eigenstates are given as
\bea
|\bl \rangle &=& \frac{\,1\,}{\,\sqrt{\,2\,}\,}
\left[ |\bo \rangle - \left( 1-2\, \varepsilon_{\rm B} \right)
e^{-\, i\, \varphi_M} |\bob \rangle \right] \nonumber \\
|\bh \rangle &=& \frac{\,1\,}{\,\sqrt{\,2\,}\,}
\left[ |\bo \rangle + \left( 1-2\, \varepsilon_{\rm B} \right) e^{-\,
i\, \varphi_M} |\bob \rangle \right]\nonumber
\eea
where $\bl$ and $\bh$ denote the light and heavy neutral B-mesons
respectively and $\varphi_M$ is the phase of $M_{12}$. The CP
violation parameter $\varepsilon_{\rm B}$ is
\be
\varepsilon_{\rm B}= - \frac{\, 1\,}{\, 4\,} \left|\frac{\, {\it
\Gamma}_{12}\, }{\, M_{12}\, } \right|
\sin \left( \delta \varphi \right)
\ee
with
\be
\delta \varphi = \varphi_M - \varphi_{\it \Gamma} + \pi
\ee
as in the kaon system. We use $\Delta {\it \Gamma}\ll \Delta
m$, i.e. $| {\it \Gamma}_{12}| \ll |M_{12} |$ which is different from
the kaon system. Note that $\varepsilon_{\rm B}$
is real which also differs from the kaon system. The standard
model predicts $\varepsilon_{\rm B}\approx {\cal O}(10^{-3})$.

In the limit of CP conservation, we have $\varphi_M=
-\theta _{\rm CP}+\pi$ {\bf and} $\varepsilon_{\rm B}=0$ so that
\be
CP |\bl \rangle = +|\bl \rangle~~{\rm and}~~
CP |\bh \rangle = -|\bh \rangle
\ee
where we assumed that the $B$ parameter in the calculation of the box
diagram is positive \cite{nakada1}. In this case, the $CP=+1$ state
is lighter than the $CP=-1$ state which is identical to the kaon
system.

The CP violation parameter for a decay into a CP eigenstate ``f'' with
$CP=+1$ is then defined as
\ben
\eta_{\rm f}= \frac{\, \langle {\rm f} | H_{\rm weak}|\bh \rangle \, }
{\, \langle {\rm f} | H_{\rm weak}|\bl \rangle \, }~.
\label{bcp}\een
In the absence of CP violation in the $\bo \bob$ oscillation,
i.e. $\varepsilon_{\rm B}=0$, $\bl$ and $\bh$ are still not CP
eigenstates. In the absence of CP violation in the decay amplitude, we
have
\be
|\langle {\rm f} |H_{\rm weak}| \bo\rangle|= |\langle {\rm f}
|H_{\rm weak}| \bob\rangle|~.
\ee
Using these relations, equation \ref{bcp} can be written as
\ben
\eta_{\rm f}=-\,i\, \frac{\, \sin \left(2\,\varphi_{\rm f} +
\varphi_{\it M}+\overline{\theta}{}_{\rm T}-\theta_{\rm CP} \right) \,
} {\, 1- \cos\left(2\,\varphi_{\rm f} + \varphi_{\it
M}+\overline{\theta}{}_{\rm T}-\theta_{\rm CP} \right) \, }
\label{CPBstand}\een
which is pure imaginary, where $\varphi_{\rm f}$ is the phase of the
$\bo
\rightarrow$f decay amplitude. Note that this corresponds more or less
to the third term in equation \ref{eta0} in the case of the kaon
system, i.e. CP violation due to interference between decay and
oscillation. In the kaon system, the phase of ${\it
\Gamma}_{12}$ given by the unitarity relation is dominated by the
$2\pi (I=0)$ decay amplitude. In the B-meson system, there exists no
dominant decay amplitude and
$\varphi_M$ (in the absence of CP violation in the oscillation this is
identical to $\varphi_{\it \Gamma}+\pi$) can be very different from
$- 2\,\varphi_{\rm f}-\overline{\theta}{}_{\rm T}+\theta_{\rm
CP}$, thus introducing a large CP violation in $\eta_{\rm f}$. In the
kaon system, this is typically $\Im (\eta _{I=0}) \approx {\cal
O}(10^{-3})$.

In the B-meson system, the standard model can make a precise
prediction for the CP violation given in equation \ref{CPBstand}.
Since $M_{12}$ is dominated by the top quark in the box diagram,
$\varphi_M$ is governed by the elements $V_{\rm tb}$ and $V_{\rm
td}$ of the Cabibbo-Kobayashi-Maskawa (CKM) \cite{CKM} quark
mixing matrix. In the decays where only one quark diagram contributes,
the phase of the decay amplitude is also given by the CKM-matrix
elements. Well known final states are J/$\psi\, \ks$  and $\pi^+
\pi^-$. Once the CKM matrix elements $V_{\rm cb}$, $V_{\rm ub}$ and
$V_{\rm td}$ become known, the CP violation parameters for these two
channels can be predicted with the help of the unitarity of the CKM
matrix
\cite{schub}.

In the B-meson system (also in the D-meson system), some final states
such as D$^+\, \pi^-$ can be produced from both $\bo$
and $\bob$ decays although they are not CP eigenstates. The CP conjugated state
D$^-\,
\pi^+$ is also the decay final state of $\bo$ and $\bob$. In this
case, the same formalism used to describe the semileptonic decay of
kaons including a possible violation of the
$\Delta Q = \Delta S$ rule can be applied.

CP violation can be
studied by comparing an initially pure $\bo$ decaying into D$^+\,
\pi^-$ with an initially pure $\bob$ decaying into D$^-\, \pi^+$. The
imaginary part of the phase convention invariant parameter
\be
x=- \frac{\, \langle {\rm D}^- \, \pi^+| H_{\rm weak} | \bob \rangle
\,}{\, \langle {\rm D}^- \, \pi^+| H_{\rm weak} | \bo \rangle
\,} e^{ i\, \varphi_{\rm M}}= - \frac{\, \langle {\rm D}^+ \,
\pi^-| H_{\rm weak} | \bo \rangle
\,}{\, \langle {\rm D}^+ \, \pi^-| H_{\rm weak} | \bob \rangle
\,} e^{ i\, \varphi_{\rm M}}
\ee
generates CP violation (see $A_{\rm CPT}$ and $A_{\rm T}$ in the kaon
system), by assuming no CP violation in decay amplitudes and
in oscillations. Note that
$|x|\ll 1$ since the decay amplitudes for $\bo \rightarrow {\rm D}^+
\pi^-$ and $\bob \rightarrow {\rm D}^- \pi^+$ are proportional to
$|V_{\rm ub}|\stimes | V_{\rm cd}|$ which is very small.

Since many decay modes are available for the B-meson system, CP
violation is expected in many different final states and could be
large. This allows us to make a consistency test by checking whether
all the observed CP violation follows the pattern predicted by the
standard model. This is essential, since other models of CP
violation can also generate large CP violation in B-meson decays
although with different patterns.

The branching ratios for the interesting
decay modes are all small requiring high statistics. More details on
various strategies to study CP violation can be found in \cite{ali}.

\subsection{Experimental Considerations}
Data on b-hadrons are so far dominated by results from
e$^+$e$^-$ machines running at the $\Upsilon$(4S)
resonance and at $Z^0$. Although high energy hadron
machines have a promising potential due to the large number of produced
b-hadrons, there is still much work needed to prove
experimental capabilities to reconstruct final states without
J$/\psi$.

The observation of CP violation, even in the most promising channel
B$\rightarrow {\rm J}/\psi \, \ks$ requires a new generation of
e$^+$e$^-$ machine in order to reach the necessary luminosities with
an asymmetric beam energy configuration when working at
$\Upsilon$(4S). Details can be found in numerous B-meson
factory proposals \cite{bfactory}.

The rate estimates in these studies should be quite reliable
due to the small background and the long experience in doing
experiments with B-mesons using such machines. The most crucial point
is whether an e$^+$e$^-$ collider working with an
asymmetric beam energy configuration can achieve the required
luminosity of ${\cal L}\ge 3\stimes 10^{33}\, {\rm cm}^{-2} \, {\rm
s}^{-1}$.

A B-meson factory working with the designed luminosity can exploit
physics requiring $10^8$ B-$\overline{\rm B}$ pairs. According to the
most resent standard model predictions \cite{schub}, CP violation
in the B$\rightarrow {\rm J}/\psi
\, \ks$ decay is within reach with such a statistics.
Its clean environment would also help to study final states of more
complex nature such as ${\rm D}^0\, \overline{\rm D}{}^0$ and $\psi'
\, \ks$.

CP violation study involving ${\rm b}\rightarrow {\rm u} + {\rm W}^-$
decays, such as the decay final state $\pi^+ \pi^-$, needs significantly
more B-mesons. Current standard model predictions \cite{schub} do not
exclude the possibility that CP violation is very small for this
channel. Recent CLEO results indicate a branching ratio of $\sim
10^{-5}$ for this decay mode \cite{cleo}.

CP violation studies in flavour specific B-meson decays provide a
clean way to eliminate the superweak model. They need also more
B-mesons. Very optimistically, the effect of CP violation could be
$\sim 10\%$ for B$\rightarrow {\rm K}^* {\rm K}$ with a branching
ratio of $\sim 10^{-5}$ \cite{wyler1}. This already shows that more
than $10^9$ B-mesons are needed.

CP violation experiments at hadron machines may become variable
if they can fully exploit the large number of produced B-mesons
\cite{hadronb}. Various projects are being considered based on both
existing machines such as Tevatron at FNAL and HERA at DESY and future
machines such as RHIC at BNL, LHC and SSC \cite{hadronb}.

Unlike for e$^+$e$^-$ B-meson factories, here the issue is the
detector capability. The high interaction rate (1 to 50 MHz depending
on $\sqrt{s}$) necessary to produce a sufficient number of B-mesons of
$> 10^{10}$ can be achieved with little problems. Whether trigger
and detector components can cope with such a high rate must be
carefully looked at. Due to high background, rate estimates will never
be certain. One can hardly generate enough simulated minimum bias
events to study the background. Since we are looking for events with
branching ratios of $\sim 10^{-5}$, any tail in various distributions
can influence the background estimation. Such a tail cannot be
reproduced with a simulation. To design an ultimate CP experiment at
a hadron machine may need more experiences from existing hadron
machines.

\section{T Violation}
There are experiments addressing the violation of T invariance
without referring to C (or CP). A typical example is the
electric dipole moment of the neutron \cite{chang} and electron
\cite{suzuki}. There are other experiments such as the scattering of
polarised protons where a triple vector product in the final state can
be used to test T violation. The standard model either does not
produce any T violation or produces very little signal and present
experiments are not sensitive enough. However, physics beyond the
standard model can produce a substantial T violation signal
\cite{beyond} which can be tested by these experiments.

\section{Conclusions}
Present experimental efforts to solve
the mystery of CP violation are large and will remain so.
We are all eager to know whether CP violation can be fully
accommodated in the standard model. The first experimental step is to
observe CP violation in the decay amplitude. This would exclude the
superweak model. Then, we have to make a quantitative comparison
between the observation and the standard model prediction.

Experiments with kaons will still dominate the field of CP
violation for some time. However, the smallness of CP
violation in the decay amplitude may forbid any discovery. Theoretical
difficulties related to low energy QCD may further hinder
quantitative comparisons.

In the long therm, the best place to make a systematic study is in the
B-meson system. CP violation is expected in many different decay
channels and reliable standard model predictions can be made for some
of the decay modes once all four parameters of the
Cabibbo-Kobayashi-Maskawa quark mixing matrix are obtained.

CP violation effects in the weak decay of the kaon are in
fact not that small compared with CP violation in strong interactions
which is naturally expected in the standard model but not observed
so far. It may be indeed that CP violation is a reflection of physics
at a much higher energy scale.

\subsection*{Acknowledgement}
All of my colleagues from the CPLEAR collaboration are especially
thanked for the exciting work we have done
together. I would like to thank D.~Cundy, R.~D.~Schaffer, A.~Wagner
and H.~Wahl from NA31, G.~Thomson from E621 and B.~Winstein and
L.~K.~Gibbons from E731 who patiently
educated me to understand their experiments. Continuous discussions
with various people, in particular with I.~I.~Bigi, G.-M.~G\'erard,
A.~Pich, K.-R.~Schubert and D.~Wyler are most appreciated and the
attempt to avoid phase conventions was strongly motivated by the
discussion with B.~Winstein. Finally, I would like to thank the
organisers of this conference for the pleasant stay at the Cornell
University and L.~K.~Gibbons who helped me as a scientific secretary
and my fellow speaker A.~I.~Sanda for discussions. K.~Gabathuler is
acknowledged for making comments and corrections to this manuscript.

\end{document}